\documentclass[12pt]{spieman}  % 12pt font required by SPIE;
\usepackage{amsmath,amsfonts,amssymb}
\usepackage{graphicx}
\usepackage{setspace}
\usepackage{tocloft}
\usepackage{xspace}
\usepackage{lineno}
\def\st{$^\text{st}$\xspace}
\def\nd{$^\text{nd}$\xspace}
\usepackage{shuffle}
\def\perio{\rotatebox[origin=c]{180}{$\shuffle$}}
\def\tur{\text{tur}}
\def\res{\text{res}}
\def\inter{\text{is}}
\def\eld{\text{d}}

\graphicspath{{Figures/}}
\title{Cascade adaptive optics: contrast performance analysis of a two-stage controller by numerical simulations}

\author[a,b,*]{Nelly Cerpa-Urra}
\author[a]{Markus Kasper}
\author[b]{Caroline Kulcsár}
\author[b]{Henri-François Raynaud}
\author[a]{Cedric Taïssir Heritier}
\affil[a]{European Southern Observatory, Karl-Schwarzschild Str. 2, Garching bei München,
Germany, 85748}
\affil[b]{Université Paris-Saclay, Institut d'Optique Graduate School, CNRS, Laboratoire Charles Fabry, 91127, Palaiseau, France}

\cftpagenumbersoff{figure}
\cftpagenumbersoff{table} 
\begin{document} 
\maketitle

\begin{abstract}
The contrast performance of current eXtreme Adaptive Optics (XAO) systems can be improved by adding a second AO correction stage featuring its own wavefront sensor, deformable mirror and real-time controller. We develop a dynamical model for such a cascade AO (CAO) system with two stages each controlled by a standard integrator, and study its control properties. We study how such a configuration can improve an existing system without modifying the first stage. We analyze the CAO architecture in general and show how part of the disturbance is transferred from low to high temporal frequencies with a nefarious effect of the 2\nd stage integrator overshoot, and suggest possible ways to mitigate this.
We also carry out numerical simulations of the particular case of a 1\st stage AO using a Shack-Hartmann wavefront sensor and a 2\nd stage AO with a smaller deformable mirror running at a higher framerate to reduce temporal error. In this case, we demonstrate that the 2\nd stage improves imaging contrast by one order of magnitude and shortens the decorrelation time of atmospheric turbulence speckles by even a greater factor. The results show that CAO presents a promising and relatively simple way to upgrade some existing XAO systems and achieve improved imaging contrasts fostering a large number of science case including the direct imaging of Exoplanets.

\end{abstract}

% Include a list of up to six keywords after the abstract
\keywords{Adaptive Optics, XAO, CAO, Speckle lifetime}

% Include email contact information for corresponding author
{\noindent \footnotesize\textbf{*}Nelly Cerpa-Urra,  \linkable{ncerpaur@eso.org} }

\begin{spacing}{2}   % use double spacing for rest of manuscript

\section{Introduction}
\label{sec:Intro}
Since the first detection of an exoplanet around a main sequence star more than 20 years ago \cite{mayor_jupiter-mass_1995}, the hunt for exoplanets has been more than prolific. Thousands of exoplanets have been detected (http://exoplanets.org/) using mostly radial velocity and transit techniques, and providing valuable information on a number of basic planet parameters such as orbit, mass, size and density. High-contrast imaging (HCI) with adaptive optics (AO), however, provides direct images of exoplanets that can be analyzed spectroscopically to characterize their atmospheres. HCI aims at reducing the host star's light flux at the location of the exoplanet, thereby minimizing the photon noise, and maximizing the detection sensitivity. It potentially reduces the required observing time to detect a planet from orbital period(s) for the indirect methods, to just a few nights or even a few hours (depending on the exoplanet's apparent flux and the measurement noise). Although more than $99\%$ of the planets discovered so far have been found indirectly, HCI led to the discovery of several young giant planets at relatively large orbital separations \cite{marois_direct_2008,marois_images_2010,lagrange_giant_2010,rameau_discovery_2013,bailey_hd_2014,macintosh_discovery_2015,chauvin_discovery_2017}, but extending the search space to lower mass and older exoplanets at smaller orbital separations has proved to be a challenge. In particular, HCI requires bigger telescopes and improved technologies (AO, coronagraphy, instrumentation) to boost contrast sensitivity at very small
angular separations. 

For optical and near-IR HCI, the AO-corrected residual halo stellar flux is the main source of measurement noise \cite{otten_direct_2021}. In order to obtain great contrast sensitivity for exoplanet imaging at small angular separations, it is therefore crucial to minimize this residual halo which is typically dominated by  the AO temporal delay at small angular separations\cite{guyon_limits_2005}. A straightforward approach to reduce the temporal delay would be to run the AO system faster at the expense of increased detector read-noise.

Ultra-fast AO systems for high-contrast imaging are under development at several observatories, using either a single-stage \cite{chilcote_gpi_2020}, a woofer-tweeter \cite{males_ground-based_2018,bond_adaptive_2020} or a cascade adaptive optics (CAO) system with two stages \cite{lozi_scexao_2018,boccaletti_sphere_2020,chazelas_ristretto_2020-1} approach. In the latter case, a 2\nd AO stage with its own deformable mirror (DM), wavefront sensor (WFS) and RTC is added to the instrument behind a 1\st stage AO system. This 2\nd stage only sees the residuals of the wavefront pre-flattened by the 1\st stage and can therefore employ a DM with small actuator stroke. As the scientific interest is mostly at very small angular separations, the AO correction radius \cite{perrin_structure_2003} can be small, and the number of actuators of the 2\nd stage's DM can be relatively low, leading to a compact design and moderate computational demands. These properties, and the possibility to develop and test the 2\nd stage stand-alone and retrofit it to an already existing 1\st stage, make this approach very attractive for upgrades of existing AO systems such as VLT-SPHERE \cite{beuzit_sphere_2019-1} or VLT-AOF \cite{madec_adaptive_2018}.

Besides running fast, predictive control presents another way to reduce the temporal error. Predictive controllers have been proposed in the literature in many different forms \cite{dessenne_modal_1997,gavel_toward_2003,le_roux_optimal_2004,poyneer_fourier_2007,piatrou_performance_2007,hinnen_exploiting_2007,petit_linear_2009,fraanje_fast_2010,kulcsar_minimum_2012,correia_spatio-angular_2015,guyon_adaptive_2017,gluck_model_2018,prengere_zonal-based_2020,liu_wavefront_2020}, and some on-sky tests have been performed \cite{dessenne_sky_1999, doelman_real-sky_2011, sivo_first_2014, lardiere_multi-object_2014,sinquin_-sky_2020}. With the greatly increased processing power and bandwidth of modern computers, predictive control has been recently brought back into the focus of AO engineering with integration in operational systems \cite{petit_sphere_2014, poyneer_performance_2016, correia_modeling_2017, males_ground-based_2018-1}.

In this work, we rather follow a simple approach and propose a CAO system with two stages, where each stage is controlled by a classical integrator. We introduce the generic CAO in Sec.~\ref{sec:principle} and study its temporal and control properties. In section 3, we present numerical simulations of a particular CAO case for an 8-m class telescope. The 1\st stage is assumed to feature a Shack-Hartmann WFS and a DM controlling about 800 modes and running at 1 kHz frame rate while the faster 4 kHz 2\nd stage features a Pyramid WFS \cite{ragazzoni_sensitivity_1999} and controls about 200 modes. This case roughly represents the considered upgrades of some VLT AO systems mentioned above.

We also analyze ways to optimize the integrator gains for both stages and how to best split the light between them. We demonstrate an improved low frequency rejection by CAO which also reduces the lifetime of atmospheric residual speckles and their noise contribution in long exposures. A comprehensive discussion of these results is provided in Sec.~\ref{sec:speckle}.

\section{Two-stage CAO system: principle and control analysis}
\label{sec:principle}
\subsection{Principle and general hypotheses}

\begin{figure}[ht]
   \begin{center}
   \begin{tabular}{c} 
   \includegraphics[height=5cm]{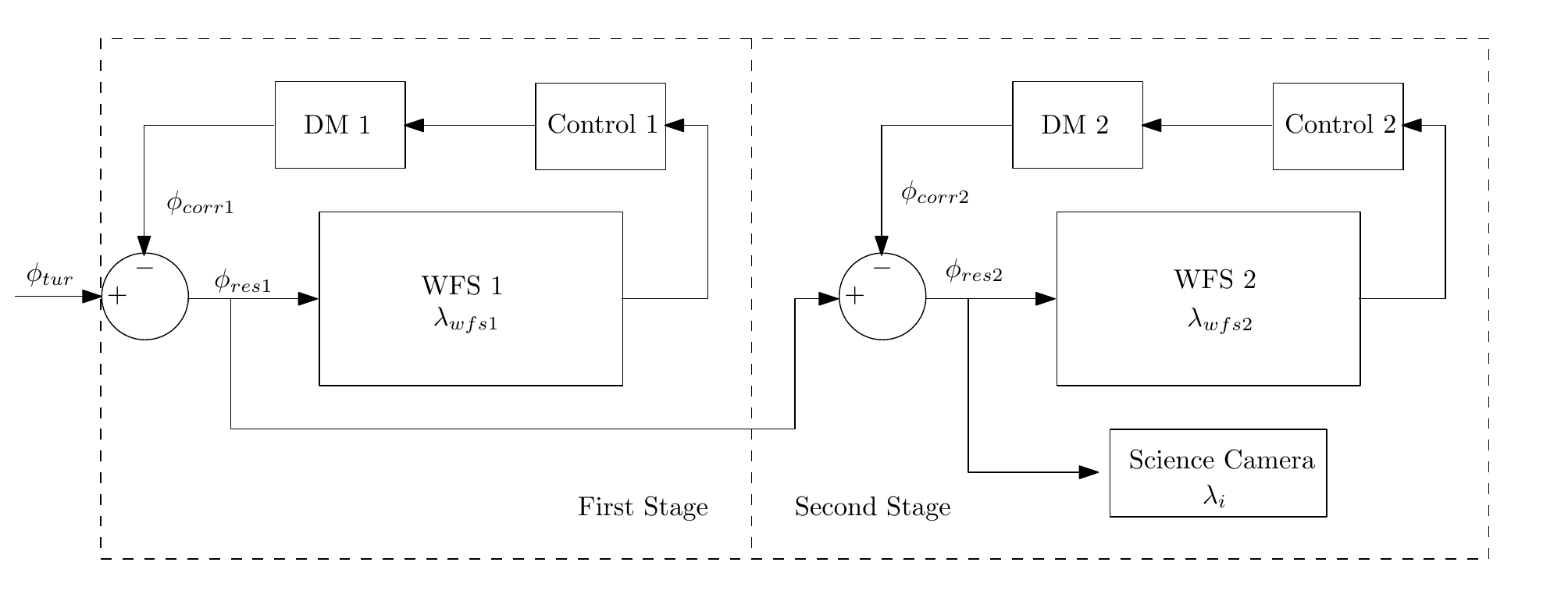}
   \end{tabular}
   \end{center}
   \caption[Two-stage CAO system architecture]
   { \label{fig:2stg}
Two-stage CAO system architecture}
\end{figure}

The principle of the two-stage CAO system is illustrated in Fig.~\ref{fig:2stg}: a 1\st stage corrects for the incoming turbulent phase, producing a residual phase that enters a faster 2\nd stage. The residual phase at the output of this 2\nd stage is sent to the science camera. The wavelengths of the two WFSs may be different, and may be also different from that of the science camera. The nature of each WFS and their dimensions will be detailed when addressing performance evaluation in Sec.~\ref{sec:perfo}.

The advantages of this kind of system can be summed up in three main points: first, the 2\nd stage can be connected to any other already existing system; this can be especially helpful by reducing downtime of existing instruments and improve implementation time. Second, a CAO system reduces the stroke needed by the 2\nd stage DM to correct for low order aberrations. And at last, the 2\nd stage system can be designed and installed without modifying the 1\st stage RTC, which is beneficial when the budget available for an upgrade is not sufficient for a complete overhaul.

Note that contrarily to standard CAO control, the 2\nd stage (inner loop) is not seen by the 1\st stage (outer loop), so that the stability of the inner and outer loops are sufficient to guarantee the stability of the CAO feedback. Each loop is controlled by an integral action controller (or integrator). Also, we will consider in the following that each WFS integrates the flux over one frame, $T_1$ for the 1\st stage, $T_2$ for the 2\nd stage, and that the two loop frequencies $F_1=1/T_1$ and $F_2=1/T_2$ are such that the 2\nd stage frequency is a multiple of the 1\st stage frequency, that is $F_2=nF_1$, $n>1,\,n\in\mathbb{N}$. Also, we suppose that each loop suffers a standard two-frame delay.

As each loop is controlled by an integrator, two natural questions arise: does this two-stage CAO system behave like a double integrator in terms of rejection? And how to characterize the rejection for such a system with two different frame rates? The purpose of the next section is to address these questions thanks to a modal analysis.

\subsection{Modal control analysis}
\label{sec:control}

In a standard AO loop, it is commonplace to analyse rejection by computing the frequency-domain response of the closed-loop controlled system. Once the frequency responses of the rejection transfer function (RTF) and noise propagation transfer function have been calculated, the effects on any incoming second-order stationary stochastic process with known power spectral density can be evaluated, allowing in particular to predict the expected value of the residual phase variance -- see, Ref.~\cite{juvenal_linear_2018} for the general case of linear controllers. These calculations rely on the hypothesis that the AO system is linear and time-invariant, so that to a given input frequency corresponds an output at the same frequency, albeit with possibly a different amplitude and nonzero phase shift.

This is no more true in the case of a two-stage CAO system featuring two different sampling frequencies, as the system looses its time invariance property. Take the case of an incoming disturbance which temporal spectrum contains energy between $F_1/2$ and $F_2/2$: this disturbance will be aliased by the 1\st stage (with an attenuation due to the averaging by the WFS) at a frequency below $F_1/2$. In addition, the frequency range $[-F_1/2,F_1/2]$ will be periodized at period $F_1$, which has to be accounted for at the 2\nd stage level. Hence, one particular frequency of the turbulent phase entering the whole two-stage system will produce several frequencies at the output of the system, so that the linearity in the frequency domain is lost: the rejection transfer function cannot be evaluated anymore as a point-by-point ratio of the output and input spectra.

We propose to analyse the rejection and noise propagation produced by this system thanks to a modal decomposition of the turbulent phase, leading to a simpler scalar temporal and frequency analysis on a single mode. All the simulations are carried out using {\sc Matlab-Simulink} with DMs and WFSs having unitary gains.

\begin{figure}[htbp]
   \begin{center}
   \begin{tabular}{c}
   \includegraphics[height=5cm]{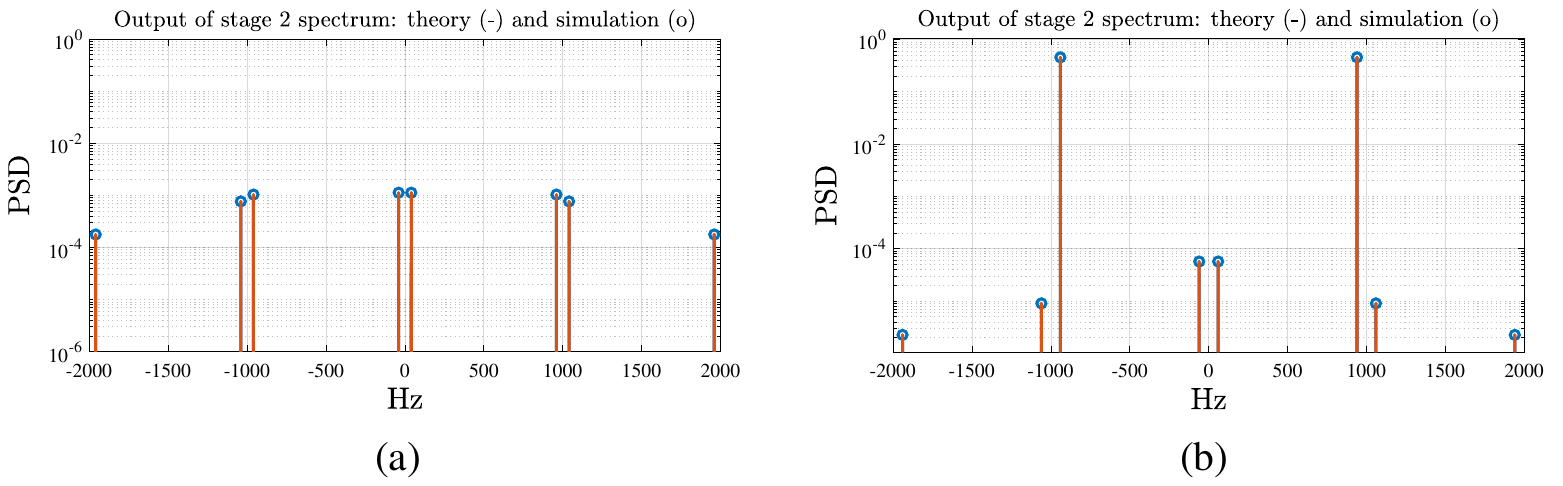}
   \end{tabular}
   \end{center}
   \caption{Illustration of spectrum periodization for an input sinusoidal signal at $f_0=40$ Hz (left) and of aliasing and periodization for an input sinusoidal signal at $f_0=940$ Hz (right). Loop frequencies are $F_1=1$ kHz and $F_2=4$ kHz. The red lines correspond to the theoretical calculations, the blue circles to the empirical power spectral density computed from the simulation data by non-averaged periodogram. The y-axis is in arbitrary units.}
 \label{fig:sinus}
\end{figure}

To illustrate the multi-rate effect, let us take a pure sine wave with frequency $f_0<F_1/2$ entering the noise-free CAO system. The 1\st stage produces a correction signal at loop frequency $F_1$, and the continuous-time residual is averaged and up-sampled by the 2\nd stage at loop frequency $F_2$. The frequency support $\{-f_0,+f_0\}$ of our initial disturbance spectrum will thus be modified into a support of the form $\{\pm f_0+mF_1\}$, $m\in \mathbb{Z}$. Figure \ref{fig:sinus}-(a) shows the spectrum of the resulting signal at the output of the 2\nd stage over the range $[-F_2/2,+F_2/2]$ for a pure sinusoidal signal of amplitude 1 and frequency $f_0=40$ Hz and loop frequencies $F_1=1$ kHz and $F_2=4$ kHz. The relative attenuation of the peaks at high frequency is due to the averaging filter convolution over $T_2$ which produces a sinc in the frequency domain. When $f_0>F_1/2$, the resulting frequencies will be of the same form $\{\pm f_0+mF_1\}$, $m\in \mathbb{Z}$, but only the aliased part of the signal will be corrected by the two stages, whereas the non-aliased part will only be corrected by the 2\nd stage. This is illustrated in Fig.~\ref{fig:sinus}-(b), and explains the higher value at $\pm f_0$. It is thus clear that the theoretical evaluation of the rejection for any given spectrum needs to distinguish what is rejected by both stages and what is only rejected by the 2\nd one.

To make this distinction, one can notice that any continuous-time signal $\phi^{\tur}$ can simply be decomposed under the form
\begin{equation} \label{eq:decomp}
\phi^{\tur}(t) = \phi^\inter(t)+ \bar{\phi}^{\tur}_k \,\,\mbox{ for  }(k-1)T_1\leq t < kT_1\,,
\end{equation}
where $\phi^\inter(t)=\phi^{\tur}(t) - \bar{\phi}^{\tur}_k$ is the so-called the inter-sampling signal and $\bar{\phi}^{\tur}_k$ is defined as
\begin{equation}
\bar{\phi}^{\tur}_k=\frac{1}{T_1}\int_{(k-1)T_1}^{kT_1}\phi^{\tur}(t) \eld t \,.
\end{equation}
The inter-sampling signal $\phi^\inter$ is not affected by the 1\st stage, only $\bar{\phi}^{\tur}$ is. Therefore, the analysis can be conducted by combining three operations:
\begin{itemize}
\item compute $\bar{\phi}^{\tur}$, the signal averaged and sampled at $T_1$, to be rejected by the 1\st stage running at $F_1$,
\item compute $\phi^\inter$, the inter-sampling signal to be compensated by the 2\nd stage only at $F_2$,
\item periodize the residual spectrum of the rejection of $\bar{\phi}^\tur$ by the 1\st stage at frequency $F_1$ on $[-F_2/2,F_2/2]$, in order to obtain the 1\st stage residual $\bar{\phi}^{\res,1}$ which is to be compensated by the 2\nd stage. The periodization is due to the upsampling at frequency F2, as $\bar{\phi}^{\res,1}$ is constant over $T_1$ and thus upsampled by repeating the same value $F_2/F_1$ times (effect of the 2\nd stage zero-order hold).
\end{itemize}
The inter-sampling signal for the 2\nd stage, obtained from $\phi^\inter$ using averages on $T_2$, has not been considered as it is not affected by the 2\nd stage.

\begin{figure}[htbp]
   \begin{center}
   \includegraphics[width=0.8\textwidth]{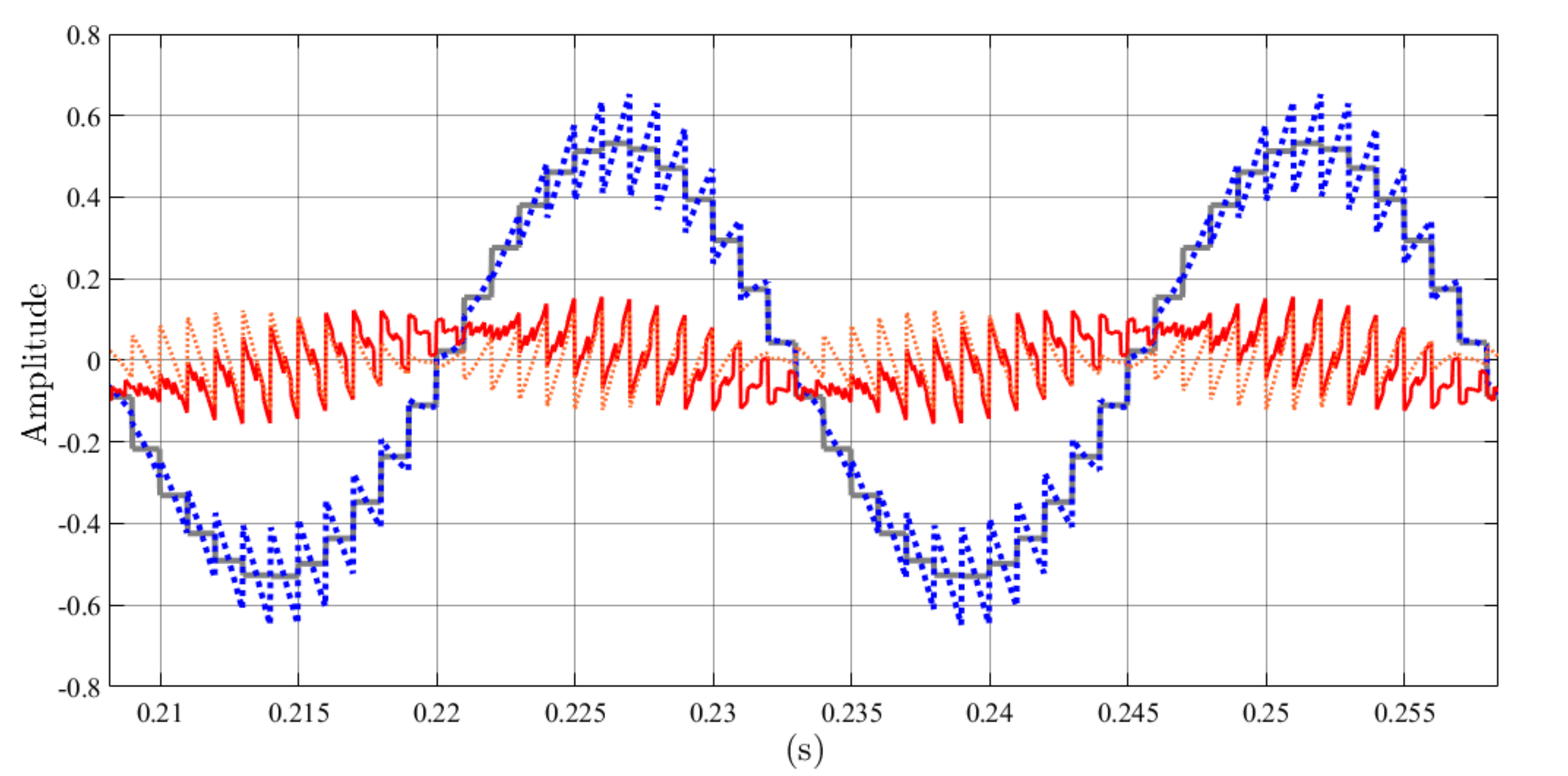}
   \end{center}
   \caption{For an input sinusoidal signal of amplitude 1 at $f_0=40$ Hz: inter-sampling signal $\phi^\inter$ (dotted orange), residual $\bar{\phi}^{\res,1}$ (plain grey), 1\st stage output $\phi^{\res,1}$ (dotted-dashed blue) and 2\nd stage residual signal $\phi^{\res,2}$ (plain red). Loop frequencies are $F_1=1$ kHz and $F_2=4$ kHz. The y-axis is in arbitrary units.}
 \label{fig:sinus_all}
\end{figure}

The complete residual signal after 1\st stage, namely $\phi^{\res,1}=\bar{\phi}^{\res,1}+ \phi^\inter$, is compensated by the 2\nd stage to give a residual signal denoted by $\phi^{\res,2}$. These various signals $\phi^\inter$, $\bar{\phi}^{\res,1}$, ${\phi}^{\res,1}$ and $\phi^{\res,2}$ are shown in Fig.~\ref{fig:sinus_all} for the 40~Hz sinusoid with amplitude 1 and for $F_1=1$ kHz and $F_2=4$ kHz. The inter-sampling signal $\phi^\inter$ (in green) plus the residual $\bar{\phi}^{\res,1}$ (in pink) gives the signal $\phi^{\res,1}$ (in blue) to be compensated by the 2\nd stage. The residual signal $\phi^{\res,2}$ (in red) is obtained at the output of the 2\nd stage. It can be noticed that the inter-sampling signal seems to have a similar energy to that of the residual $\phi^{\res,2}$.

\begin{figure}[htbp]
   \begin{center}
   \includegraphics[width=0.7\textwidth]{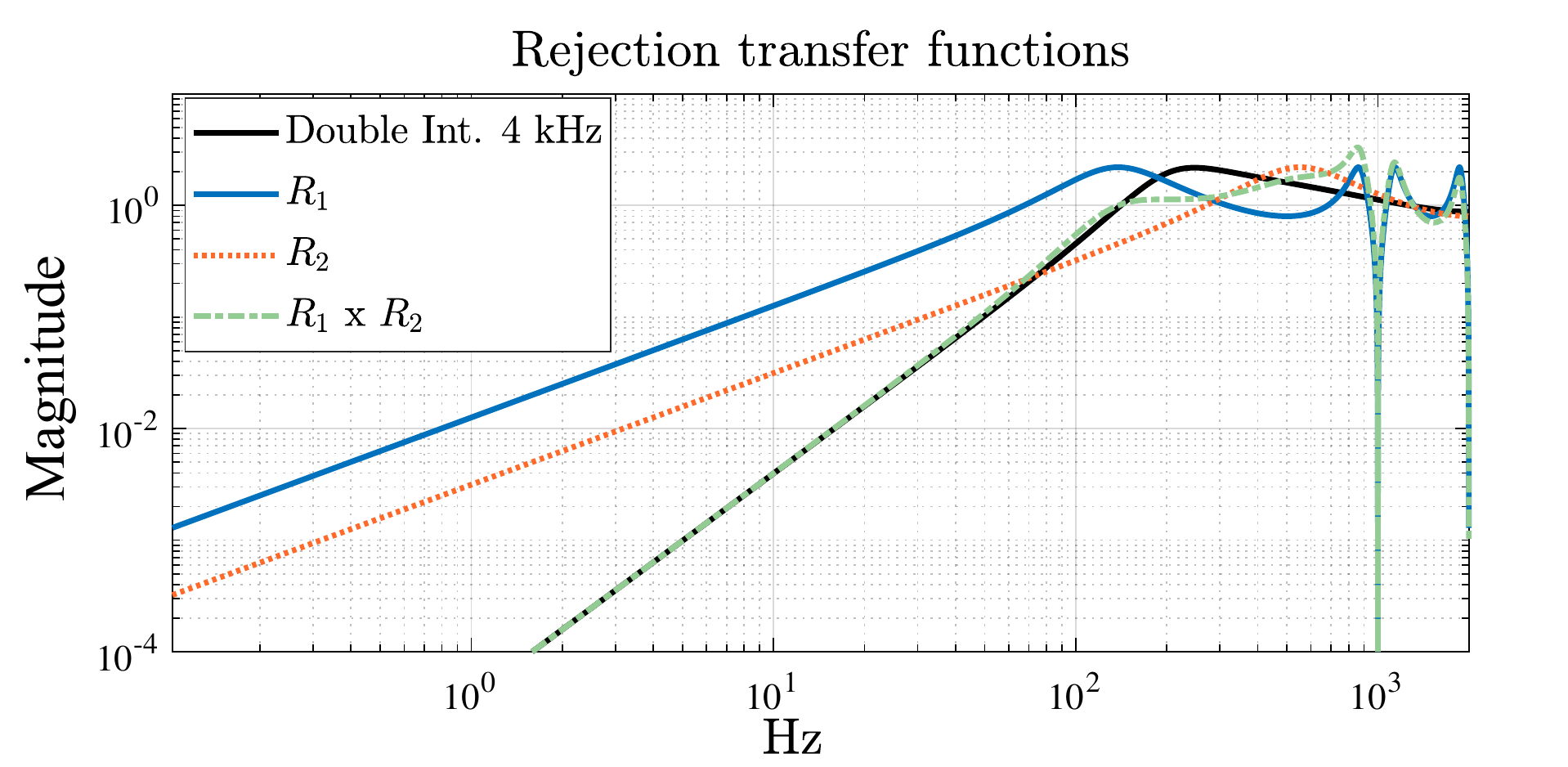}
   \end{center}
   \caption{RTFs over the frequency range $[0,\, F_2/2]$. The signal $\bar{\phi}^\tur$ generated at $F_1=1$ kHz and upsampled at $F_2=4$~kHz is rejected by $R_1\times R_2$, while the inter-sampling signal $\phi^\inter$ is only rejected by $R_2$. The double integrator RTF at $F_2=4$~kHz is in black. The y-axis is in arbitrary units.}
 \label{fig:RTFs}
\end{figure}

The rejection does not affect all these signals in the same way: if we denote by $R_1$ and $R_2$ the rejection transfer functions (RTFs) of stages 1 and 2 respectively, and considering the decomposition in Eq. \eqref{eq:decomp} and the explanations given above, the power spectral density (PSD) $S_{\res,2}$ of the residual signal at the output of the 2nd stage will be given by

\begin{equation}
S_{\phi^{\res,2}}(e^{i\omega T_2})=|R_2(e^{i\omega T_2})|^2\left[S_{\phi^\inter}(e^{i\omega T_2})+ \perio\left(| R_1|^2 S_{\bar{\phi}^\tur}\right)(e^{i\omega T_2})\right] \,,
\end{equation}
where $S_{\phi^\inter}$ and $S_{\bar{\phi}^\tur}$ are respectively the PSDs of $\phi^\inter$ and $\bar{\phi}^{\tur}$, and $\perio(G)$ is the periodized version of $G$ with period $F_1$ on the interval $[-F_2/2,+F_2/2]$. In order to evaluate the rejection and compare it with a double integrator at $F_2$, Fig.~\ref{fig:RTFs} displays the various RTFs that affect the signals passing through the system differently for $F_1=1$ kHz and $F_2=4$ kHz. The gains for the two integrators is 0.5, and the double integrator (with gain 0.0625) has been stabilized using a lead-lag term $1+\alpha(1-z^{-1})$ with $\alpha=4$ in order to limit the frequency-domain overshoot. From Fig.~\ref{fig:RTFs}, one can see that the two-stage CAO system will reject the low frequency content of a signal like a double integrator until about 100 Hz. However, the inter-sampling signal (which is only rejected by $R_2$) has a spectrum which spreads until high frequencies because of $F_1$ periodization. It will thus not be well attenuated by the 2\nd stage and may even be amplified by the overshoot of the integrator. As for high-frequency and high-energy input signals above about 300 Hz, they are unlikely to be present in an atmospheric perturbation.
The two-stage system is expected to have a better behavior in the range 180-300 Hz because it is well below the overshoot of the double integrator.

\begin{figure}[htbp]
   \begin{center}
    \includegraphics[width=1\textwidth]{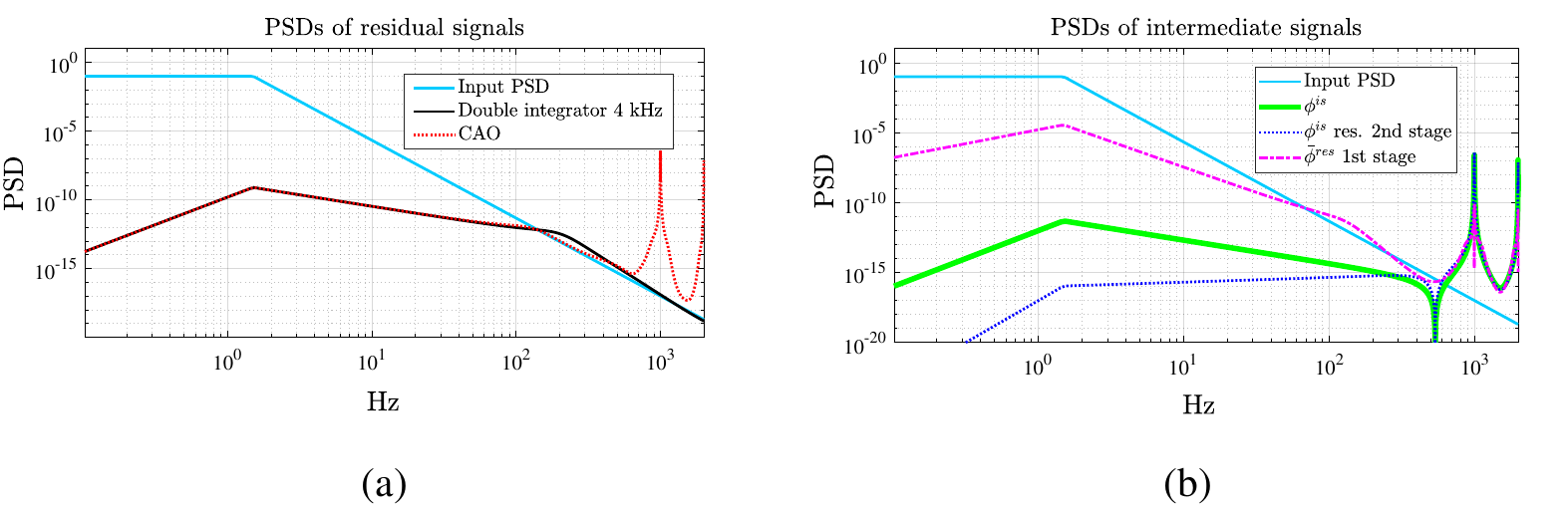}
   \end{center}
   \caption{Theoretical PSDs: on the left, schematic input PSD corresponding to a Zernike radial order $n_\text{rad}$=3, with cut-off frequency $f_c=1.5$ Hz (plain blue), residual PSD $S_{\res,2}$ at the output of the 2-stage CAO system with loop frequency $F_2=4$ kHz (dotted red) and residual PSD with a double integrator at $F_2=4$ kHz (black). On the right, from top to bottom: input PSD (light blue), PSD of residual signal $\bar{\phi}^{\res,1}$ at the output of 1\st stage with loop frequency $F_1=1$ kHz (dashed magenta), inter-sampling signal $\phi^\text{is}$ PSD (thick green) and PSD of the residual inter-sampling signal after rejection by the 2\nd stage RTF $R_2$ (dotted dark blue).}
 \label{fig:PSDs}
\end{figure}

Let us take now the example of the temporal spectrum of a Zernike mode of radial order $n_\text{rad}$=3, with cut-off frequency $f_c=0.3(n_\text{rad}+1)V/D=1.5$ Hz ($V=$ 10 m/s, $D=8$ m)\cite{Conan_1995}. The schematic PSD, 2-stage residual PSD and double integrator residual PSD are plotted in Fig.~\ref{fig:PSDs}-(a) for a case without measurement noise. The behavior of the 2-stage system is similar or better than that of the double integrator for the part of the spectrum until about 600~Hz. The inter-sampling signal produces the two peaks at high frequency, leading to a global variance that is above the double integrator. In Fig.~\ref{fig:PSDs}-(b), it can be seen that the rejection of $\bar{\phi}^{\res,1}$ is to be improved at low frequencies, which is well done by the 2\nd stage as shown with the CAO curve in Fig.~\ref{fig:PSDs}-(a). The high frequencies of the inter-sampling signal $\phi^\inter$ stay almost identical after the 2\nd stage as they are not attenuated, as noticed previously in the Fig.~\ref{fig:sinus_all}, and as it can also be seen in Fig.~\ref{fig:RTFs}. The inter-sampling signal thus dominates the global signal in terms of variance, which can be computed using the formula in \cite{kulcsar_minimum_2012}:

\begin{equation}
\text{Var}(\phi^\text{is})\stackrel{\rm a.s.}{=} { \int_{-\infty}^{+ \infty} \left( 1-|\mbox{sinc} (\pi f  T_1)|^2 \right) S_{\phi^\tur}(f)
  \eld f} \,,
\end{equation}
where $S_{\phi^\tur}(f)$ is the PSD of $\phi^\tur(t)$. For this schematic spectrum, with $\text{Var}(\phi^\tur)=3.4$ arbitrary units [arb. units] over $[-F_2/2,\, F_2/2]$, one finds $\text{Var}(\phi^\text{is})\simeq 1.4\,10^{-5}$ [arb. unit] when the variance of the residual signal at the output of stage 2 is Var$(\phi^\text{res,2}) \simeq 2\, 10^{-5}$ [arb. unit]. The double integrator is of course lower with a residual variance of about $4\, 10^{-8}$ [arb. unit].

\begin{figure}[htbp]
   \begin{center}
    \includegraphics[width=1\textwidth]{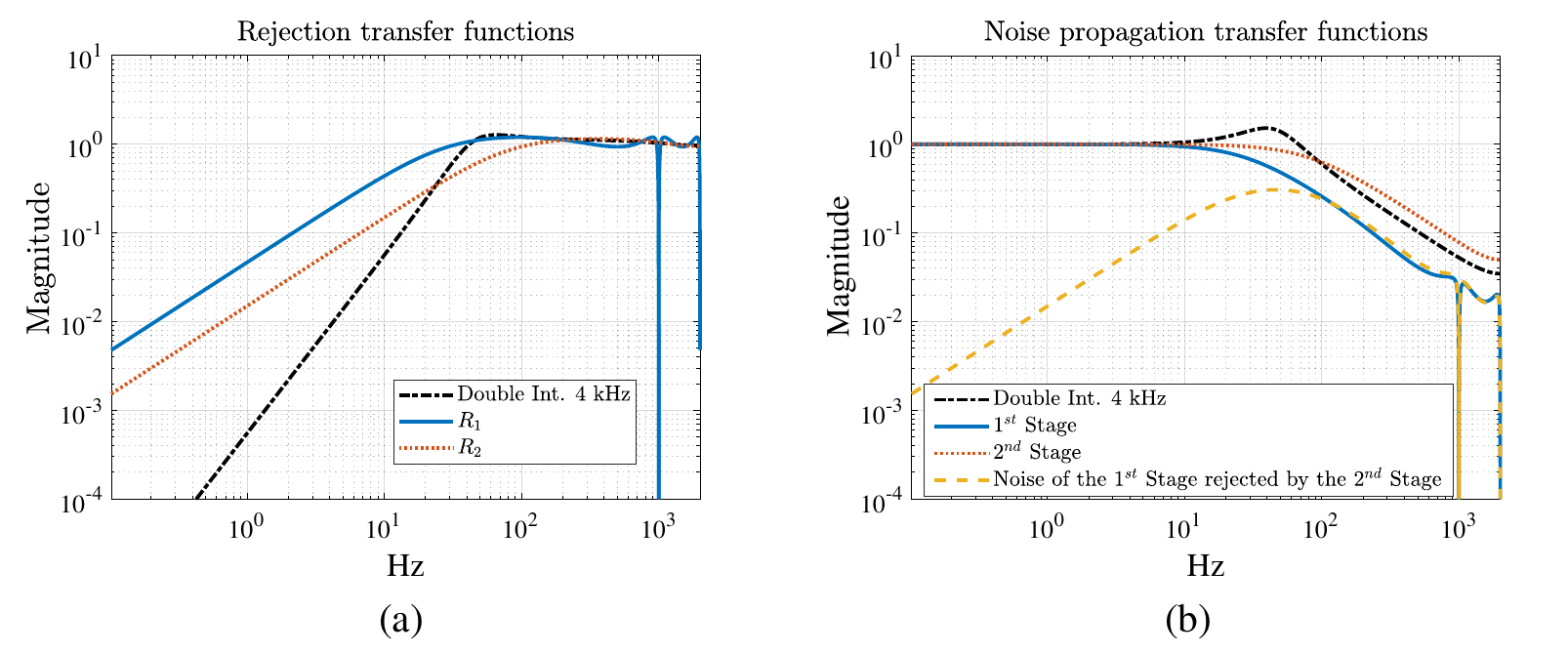}
   \end{center}
   \caption{Transfer functions. On (a), RTFs with best tuning for each regulator: double integrator (Dashed black line), 1\st stage (blue line), 2\nd stage (dotted red line). On (b), noise propagation transfer functions: double integrator (dashed black line), 1\st stage (blue line), 2\nd stage (red dotted line), 1\st stage noise propagation rejected by 2\nd stage (yellow dashed line).}
 \label{fig:NRTFs}
\end{figure}

When measurement noise is present, its propagation needs to be accounted for even in a very low noise situation, as shown in the following. The dimensioning of the 1\st and 2\nd stages leads to consider a similar noise variance for each stage, and a value of $4 10^{-4}$ [arb. units] has been chosen. This corresponds to a measurement noise variance about eight thousand times lower than that of the mode, that is a very high flux condition.
We have tuned globally the two integrator gains of the CAO and also the gain and parameter of the double integrator and lead-lag filter to obtain the best residual variances: 6.0~$10^{-5}$ [arb. units] for the CAO, and 3.9~$10^{-5}$ [arb. units] for the double integrator. The corresponding rejection transfer functions and noise propagation transfer functions are shown in Fig.~\ref{fig:NRTFs}.
On the right, the transfer function of the 1\st stage noise rejected by the 2\nd stage corresponds to the noise propagation transfer function of stage 1 multiplied by the stage 2 RTF (illustrated in dotted line on the left). It shows that the 1\st stage gain cannot be set at a too high value because of the propagated measurement noise that attacks the 2\nd stage RTF. Even in a very high flux situation, as can be seen in Fig. \ref{fig:OptiGainCheck}(a) for a magnitude of 3, the 1\st stage gain must be chosen not too high to limit noise propagation.

\begin{figure}[htbp]
   \begin{center}
    \includegraphics[width=1\textwidth]{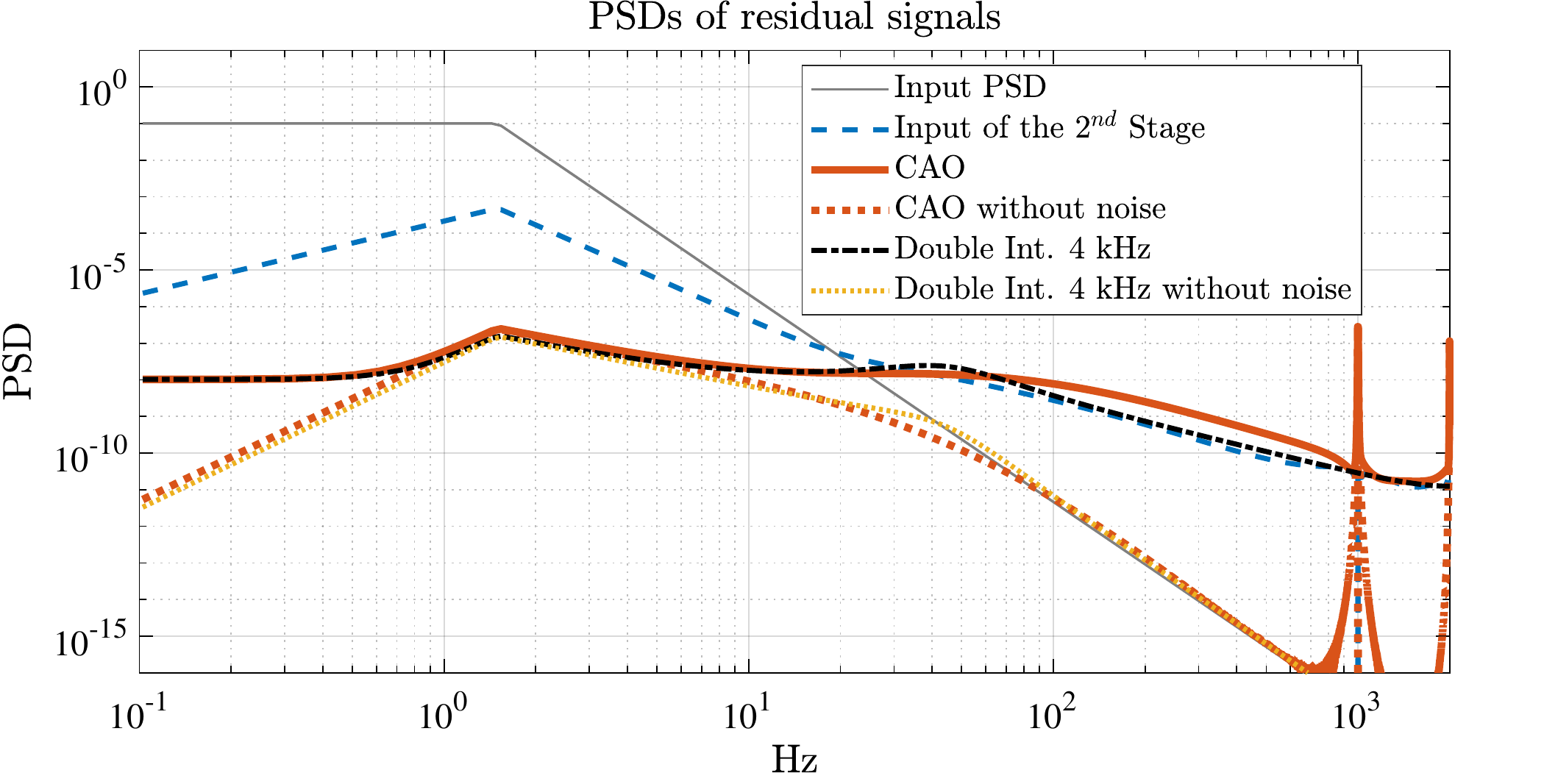}
   \end{center}
   \caption{PSDs in presence of a measurement noise variance of $4 10^{-4}$ [arb. units]: schematic modal input PSD (grey), input of the 2\nd stage including inter-sampling signal (dotted blue), residual at the output of the CAO (plain red and dashed-dotted red for the case without noise), residual for the double integrator (dashed-dotted black and dotted yellow for the case without noise). The peaks of the 2\nd stage input are at the same level than that of the CAO.}
 \label{fig:PSDsNoise}
\end{figure}

The theoretical residual phases PSDs when noise is present noise are shown in Fig. \ref{fig:PSDsNoise}. 
The variance of the inter-sampling signal, $1.4\,10^{-5}$ [arb. unit], becomes now moderate compared with the total residual variance of $6.0\,10^{-5}$ [arb. units] for the CAO. 

In summary, this two-stage CAO controller with loop frequencies $F_1=1$ kHz and $F_2=4$~kHz, when applied to standard atmospheric perturbations, will generate high frequencies (the inter-sampling signal due to the presence of the 1\st stage) from low-frequency signals. The inter-sampling signal variance depends on the turbulence strength, with high frequencies not well attenuated by the 2\nd stage because of the integrator overshoot. However, its impact is moderate in presence of measurement noise which propagation has a decisive impact on global performance. The low frequencies (until about twenty Hertz) will be on the other hand attenuated at the same level than that of a double integrator, and a better attenuation can be expected with the CAO at frequencies where the double integrator overshoots. We have also seen that the 1\st stage propagated noise enters the 2\nd stage as a disturbance, so that the value of the 1\st stage integrator gain should be chosen not too high.

The CAO will provide a better contrast than the single stage system with good correction at low angular separations (corresponding to low order modes and low spatial frequencies \cite{guyon_limits_2005}). It is clear here that a regulator designed specifically to compensate the 2\nd stage input disturbance while limiting the 2\nd stage noise propagation should further improve the global rejection, which is left for future work. The two-stage system presented here is simple to tune and its stability margins can be set separately in each loop by a proper choice of the two integrator gains. This modal control analysis has now to be completed by a performance analysis in terms of contrast and speckle lifetime, which is the purpose of the following section. 

\section{Performance Simulations}

Operation and performance of a CAO system with two stages can be studied with the help of numerical simulations. As the CAO features two wavefront sensors, we have to make assumptions about the WFS types, the framerate at which these are running, and the beam-splitting between the two. We choose the Shack-Hartmann Sensor (SHS) to drive the 1\st stage, consistent with the CAO concepts for SPHERE+ \cite{boccaletti_sphere_2020} and RISTRETTO \cite{chazelas_ristretto_2020-1} which are currently in early development phases for ESO telescopes. The 2\nd stages of these concepts (and our simulations) are driven by a unmodulated  Pyramid WFS (PWS) which provides significantly better sensitivity and wavefront sensing accuracy than the SHS \cite{ragazzoni_sensitivity_1999}. In particular, the advantage of a high-order AO system operated by a PWS is greatest near the image center which is where most Exoplanets appear and where the Exoplanet science case benefits most from a better WFS \cite{verinaud_nature_2004, guyon_limits_2005}. The 1\st stage's SHS is chosen to control about 800 Karhunen-Lo\`{e}ve modes sampled by 36x36 subapertures at 1 kHz similar to what SPHERE SAXO \cite{fusco_final_2014} and the AOF provide\cite{Madec_AOF_2018}. The 2\nd stage's PWS has a twice coarser one-dimensional sampling (18x18) and controls 200 modes, but runs four times faster (4 kHz) to efficiently reduce temporal error. These ballpark figures are consistent with what is considered for the 2\nd stage AO systems under development mentioned above.  

We explore two different options for the beam-splitting between the two stages: a) gray beam-splitting with a variable fraction of the I-band intensity distributed between the two stages, and b) dichroic beam-splitting with the 1\st stage operating at a longer wavelength in the J-band, and the 2\nd stage operating in I-band. These two beam-splitting cases are scientifically and technically motivated. The science case for SPHERE+ is focused on young stellar objects, and RISTRETTO is ultimately designed for the observations of Proxima b which is an approximately Earth-mass planet orbiting our nearest neighboring star Proxima Centauri \cite{anglada-escude_terrestrial_2016}. Temperate small planets were also found around several other very nearby stars \cite{ribas_candidate_2018,diaz_sophie_2019,zechmeister_carmenes_2019}, and many more are expected to be identified by existing and future radial velocity (RV) instruments \cite{quirrenbach_carmenes_2018,wildi_nirps_2017}. The exoplanet host stars for these science cases, either very young or very nearby, are typically of a late spectral type and emit most of their flux in the I-band or longer wavelengths. This is also where important molecular lines can be found in the planetary spectra such as the A-band of molecular Oxygen at 760 nm which is the science wavelength we consider for our analysis. We choose the PWS of the 2\nd stage to operate in I-band in order to have it as close as possible to the science wavelength and minimize chromatic residuals \cite{guyon_limits_2005}. The 1\st stage WFS could then operate in the near IR, e.g., the J-band, and the light would be split between the two by a dichroic. 

The 1\st stage could also operate in I-band in which case a gray beam-splitter would be used. This would be a technically simple solution keeping the existing 1\st stages of SPHERE and AOF, and we explore the best splitting ratio between the two stages hereafter. Table \ref{tab:System-Parameters} summarizes the observation and instrument parameters used for the numerical simulations.

\begin{table}[htbp]
\caption{System Parameters} 
\label{tab:System-Parameters}
\begin{tabular}{ll}
\hline
\textbf{Atmosphere}             &                                                       \\
$r_0$                              & $0.10$ [m] and $0.157$ [m] at $550$ [nm]                                         \\
$L_0$                            & $25$[m]                                                  \\
Fractional $r_0$                   & {[}53.28 1.45 3.5 9.57 10.83 4.37 6.58 3.71 6.71{]}\% \\
Altitude                        & {[}42 140 281 562 1125 2250 4500 9000 18000{]} m      \\
Wind Speed                      & {[}15 13 13 9 9 15 25 40 21{]} m/s                    \\
Wind Direction                  & {[}38 34 54 42 57 48 -102 -83 -77{]}$\times \pi/180$           \\ \hline
\textbf{Telescope}              &                                                       \\
Diameter                        & $8$ [m]                                                   \\

Secondary Diameter     & $1.16$ [m]                                                \\ 
\hline
\textbf{Photometric System}              &                                                       \\ 
I Band                          & Wavelength: $0.790e-6$; Bandwidth: $0.150e-6$              \\ 
J Band                          & Wavelength: $1.215e-6$; Bandwidth: $0.260e-6$              \\ \hline
\textbf{Guide-Star (Proxima Centauri)}             &                                                       \\
I-J Color Index               & $2.06$                                                   \\ 
Apparent magnitude (J)                 & $5.35$  
\\
Apparent magnitude (I)                & $7.41$\\ \hline
\textbf{Science Camera}         &                                                       \\
$\lambda_i$                      & I-band                                                \\ \hline
\textbf{1st Stage}              &                                                       \\
WFS                             & Shack-Hartmann                                        \\ 
Order WFS                           & $36 \times 36$                                                 \\ 
Control Modes                   & $800$                                                   \\ 
$n_{pix}$ Camera                        & $216 \times 216$ [pixels]                                                     \\ 
DM                              & $37 \times 37$                                                 \\ 
$\lambda_{wfs}$                   & J-band or I-band                                                \\ 
Transmission                     & $0.2$                                                   \\ 
QE                              & $0.5$                                                 \\ 
Readout Noise                   & $0.5$ [electron per pixel]                                                   \\ 
Loop Frequency $F_1$                  & 1 kHz                                                 
                                     \\ \hline

\textbf{2nd Stage}              &                                                       \\
WFS                             & Pyramid                                               \\
Modulation                      & Unmodulated                                           \\ 
Order WFS                       & $18 \times 18$                                                 \\ 
Control Modes                   & $200$                                                   \\ 
 $n_{pix}$ Camera                        & $216 \times 216$  [pixels]                                                    \\
DM                              & $19 \times 19$                                                 \\
$\lambda{wfs}$                    & I-band                                                \\
Transmission                     & $0.2^{(a)}$                                                   \\ 
QE                              & $0.5$                                                  \\ 
Readout Noise                   & $0.5$ [electron per pixel]                                                  \\ 
Loop Frequency $F_2$                 & 4 kHz                                                 
                                    \\ \hline
\end{tabular}
\footnotesize{\\$^{(a)}$The additional transmission losses for the 2nd Stage due to the slightly increased number of optical surfaces \\ is negligible and the same transmission has been assumed for both WFS.\\}
%\footnotetext{}
\end{table}

To simulate this CAO concepts, we use the AO simulation package {\sc OOMAO} \cite{conan_object-oriented_2014} running under {\sc Matlab}, and we implemented an integrated solution for simulating both stages in a single simulation run. For this, we generate the input phase at the fastest frequency (in our case, 4 kHz) and input an average of four consecutive turbulent phase screens to the 1\st stage (1 kHz) at every fourth step and update the 1\st stage DM. Then, the residual phase generated by the 1\st stage is sent as an input to the 2\nd stage at each step. This configuration allows us to take into account changes of the turbulent phase at the fast rate on the 2\nd stage. We will now describe how the two main operation parameters of the CAO, the beam-splitting ratio between the two stages and their integrator gains, were optimized.

\subsection{Optimization of integrator gains and gray beam-splitting ratio}
\label{sec:bs_gain}

Depending on the WFS incident flux, the integrator gains must be adjusted for optimum performance. For simplicity, we assume a global gain for a given stage, but note that a modal gain optimization \cite{gendron_astronomical_1994} can lead to an improved correction performance especially for faint stars. The incident flux on the WFS detector takes into account the assumed transmission to the detector listed in Table~\ref{tab:System-Parameters} and the WFS wavelength bandpass. For the dichroic beam-splitting, we assume a stellar with spectral type M5 (e.g. Proxima Centauri with I-J = 2.06) as a template red star, to calculate the flux intensities in the different bands. For example, a J = 0.94 and I = 3 star provides 15710 and 2600 photons/subaperture/frame on stage 1 and 2, respectively. For the gray beam-splitting case, five different split ratios between the 1\st stage and the 2\nd stage were simulated: 20\%/80\%, 35\%/65\%, 50\%/50\%, 65\%/35\% and 80\%/20\%. For example, an I = 3 star observed with an 80/20 beam-splitter provides 2199 and 521 photons/subaperture/frame on stage 1 and 2, respectively. Here, the relative photon flux approximately corresponds to the split ratio, because the 2\nd stage has four times bigger subapertures than the 1\st stage but runs four times faster.

\begin{figure}[ht]
   \begin{center}
   \begin{tabular}{c} 
   \includegraphics[height=7.3cm]{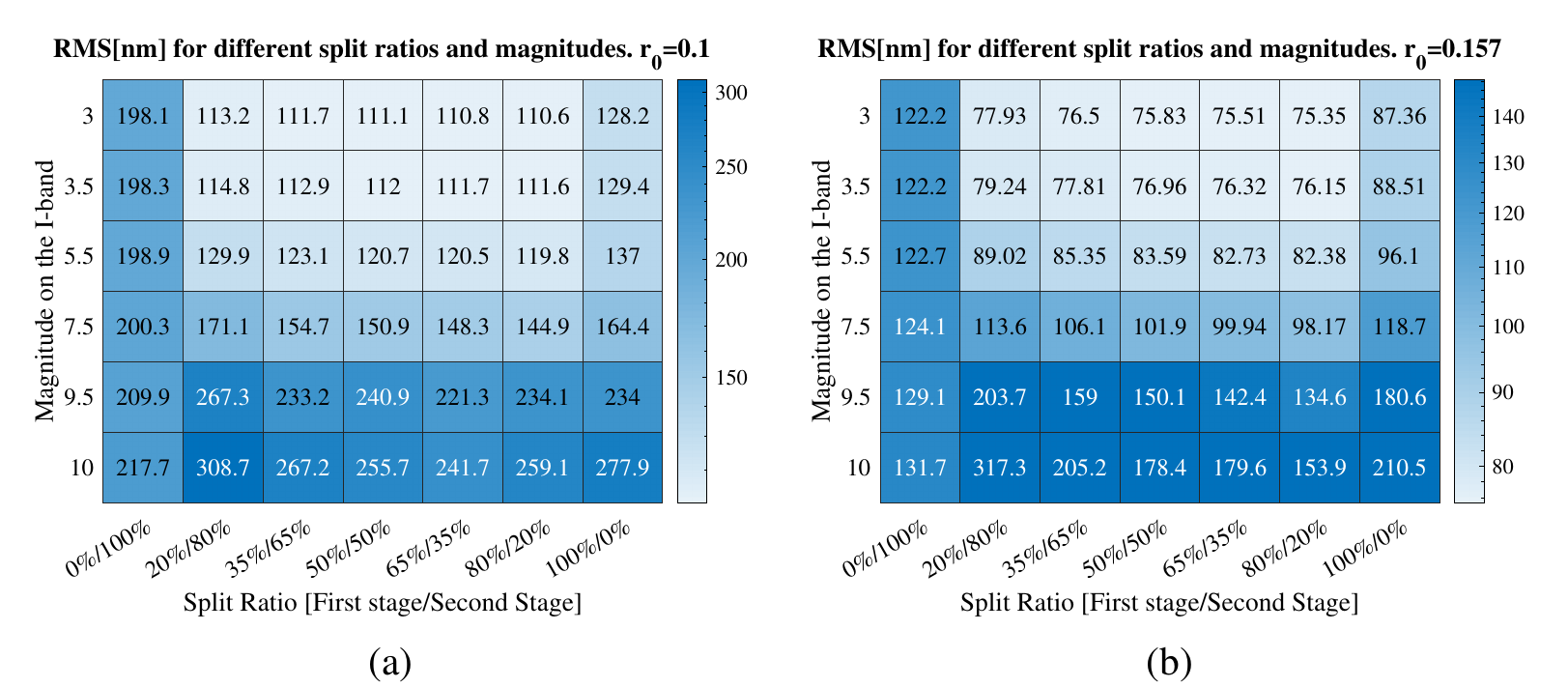}
   \end{tabular}
   \end{center}
   \caption
   { \label{fig:SplitVSMag} Wavefront error [nm] for different split-ratios (gray beam-splitter case) and magnitude. On the left side, the simulation were done using $r_0 =$ 0.1 [m] and on the right side $r_0 =$ 0.157 [m]. The darker the color, the bigger the WFE. In terms of RMS, the best overall performance is achieved using an 80\%/20\% split ratio for magnitudes up to~8 in I-band.}
\end{figure}

In order to determine the optimum gray beam-splitting ratio, we first individually optimize the integrator gains for both stages depending of the incident flux and split ratio. These results are shown in Fig.~\ref{fig:SplitVSMag}. Apart from the faintest stars, the best overall performance is obtained when 80\% of the light is sent to the 1\st stage and 20\% is sent to the 2\nd stage for both values of $r_0$ evaluated. This reflects the higher sensitivity of the PWS when compared with the SHS. The 2\nd stage maintains a high performance on significantly lower photon flux than what is required for the 1\st stage. 

\begin{figure}[ht]
   \begin{center}
   \begin{tabular}{c} 
   \includegraphics[height=10cm]{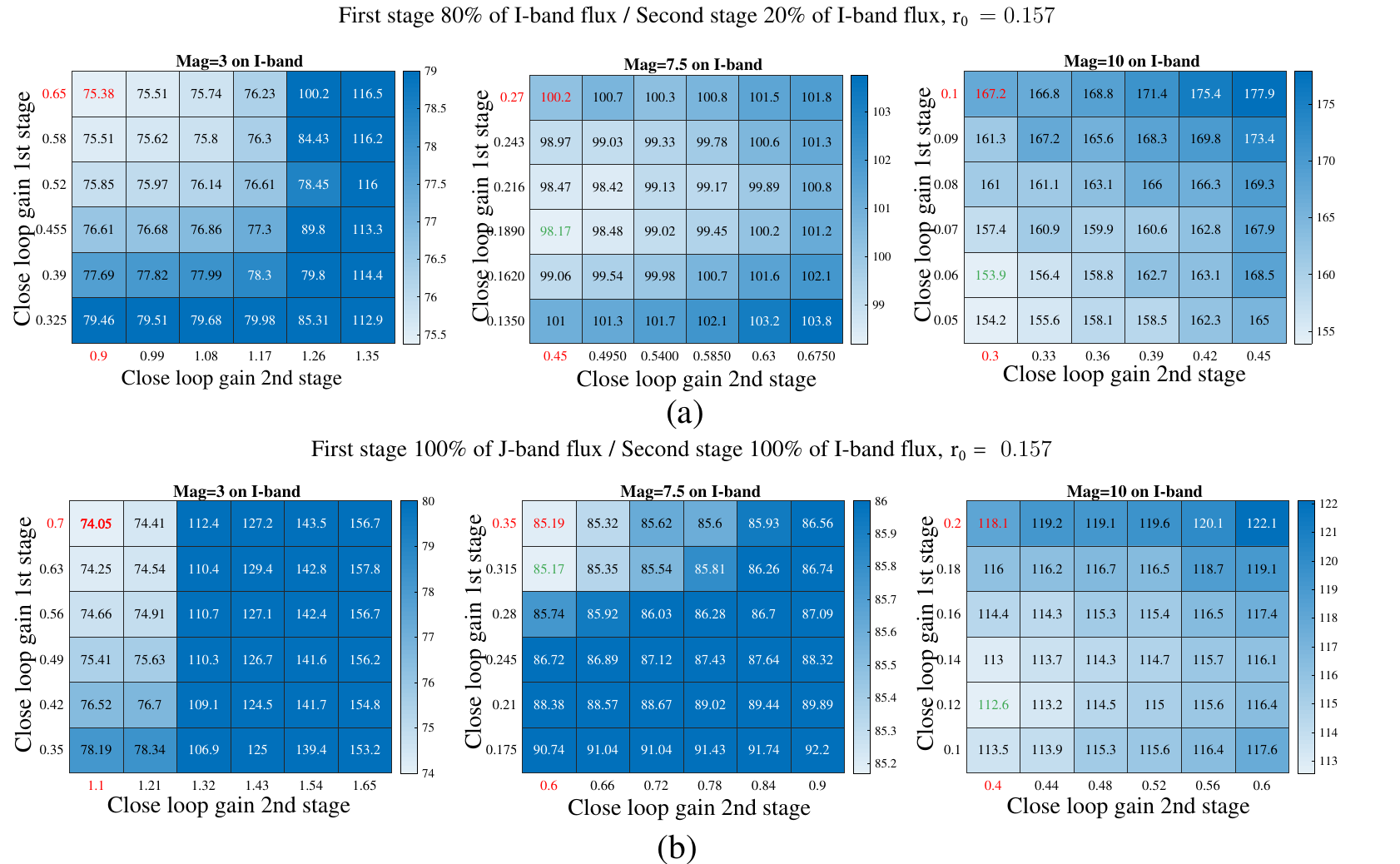}
   \end{tabular}
   \end{center}
   \caption[Residual WFE as a function of close loop gain and three different magnitudes.] 
   { \label{fig:OptiGainCheck} Residual WFE [nm RMS] as a function of close loop gain and three different magnitudes. Panel a) on the top shows the gray beam-splitting case using a 80\%/20\% split-ratio, while on panel b) on the bottom, we assume a dichroic beam-splitter.  The starting point of the joint optimization are the individually optimized gains shown in red with the corresponding WFE in the upper left corner. From there, we explored reduced stage 1 gains and increased stage 2 gains. The minimum residual WFE obtained by the joint optimization is shown in green. For the bright guide star case (I=3), the individually optimized gains were already optimum.}
\end{figure}

Then we jointly optimize the gain values of both stages for both beam-splitting cases. In order to limit the size of the already large simulation parameter grid, we only consider the 80/20 split ratio for this analysis. The rationale for the joint optimization of the gains is that the PSD of the input disturbance to the 2\nd stage is modified by the rejection transfer function (RTF) of the 1\st stage controller (as described in Sec.~\ref{sec:control}). A higher gain for the 1\st stage will lead to better low frequency rejection but more overshoot at high frequencies. It therefore shuffles energy from low to high frequencies where the correction by the 2\nd stage is less effective. Hence, a 1\st stage gain that optimizes its own residual wavefront variance may not lead to the minimum residual wavefront variance at the output of the 2\nd stage.
Figure~\ref{fig:OptiGainCheck} shows an example results of the joint gain optimization for a bright, a medium and a faint guide star, respectively. For each magnitude we diminish the 1\st stage closed-loop gain obtained by the individual optimization from a factor 1 to 0.5 in 10\% steps and we amplify the 2\nd stage closed-loop gain obtained by the individual optimization from a factor 1 to 1.5 in 10\% steps. The resulting gains for stage 1 and 2 are shown on the Y- and X-axes of the heatmaps, respectively. The results for the individual optimization of the integrator gains are shown in red, while the joint optimization results are shown in green. We see that in general, the joint gain optimization leads to a reduced 1\st stage gain with respect to the individual optimization but maintains the 2\nd stage gain. For bright magnitude = 3 stars, there is no evident improvement in performance by using the joint gain optimization probably because the high-frequency amplification mostly consists of noise, so the reduction of the fitting error of modes not controlled by the second stage is more relevant than the amplification of the low level of noise at high temporal frequencies.

\begin{figure}[ht]
   \begin{center}
   \begin{tabular}{c} 
   \includegraphics[height=4cm]{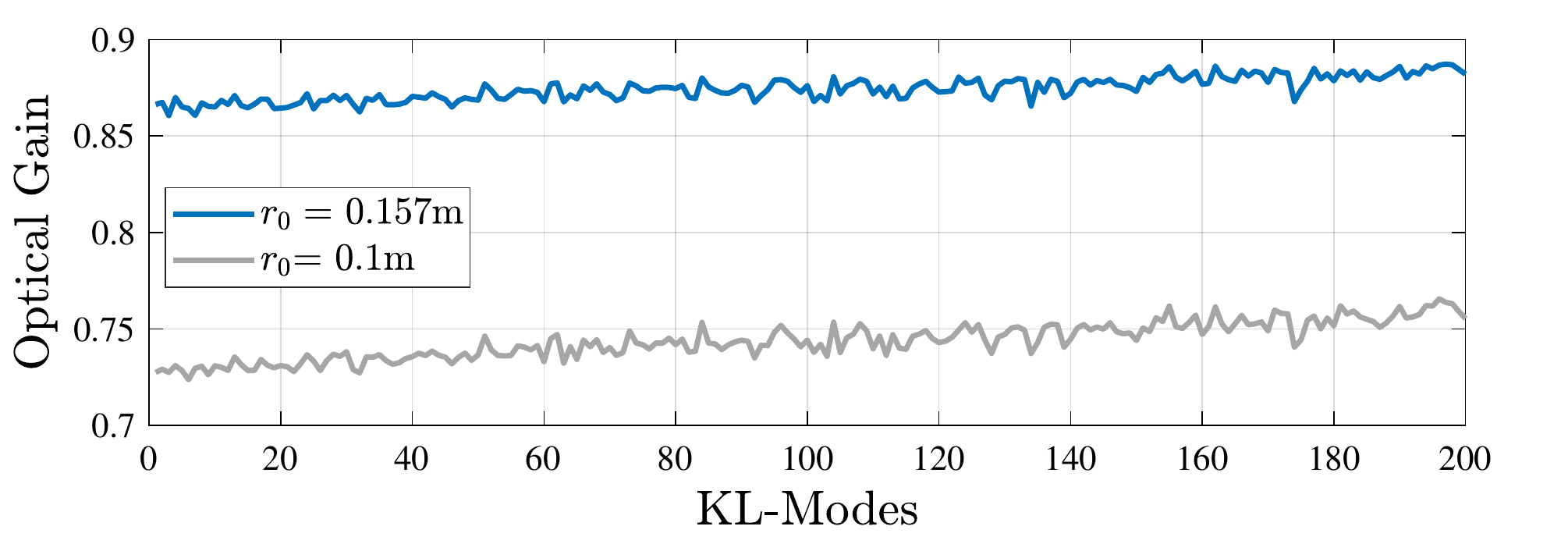}
   \end{tabular}
   \end{center}
   \caption[Optical gains for the 2\nd stage PWFS as a function of the 
   %Fried parameter $r_0$.] 
   Karhunen-Loève modes for two different values of the Fried parameter $r_0$.] 
   { \label{fig:OpticalGain} Optical gains for the 2\nd stage PWFS as a function of the Karhunen-Loève modes for two different values of the Fried parameter $r_0$.}
\end{figure}

Note that the optimum gains for the 2\nd stage include the optical gains of the PWS \cite{korkiakoski_applying_2008,deo_modal_2018} which are always smaller than one. Using a published method \cite{deo_modal_2018} we calculate optical gain values ranging between 0.75 for $r_0$ = 0.1m and 0.87 for $r_0$ = 0.157m as shown in Fig.~\ref{fig:OpticalGain}. The optical gains are relatively large because the 2\nd stage PWS only sees the residuals of the 1\st stage and therefore operates in a very small residual WFE regime. Then, the pure integrator control gain would be obtained by multiplying the 2\nd stage gain with the respective optical gain. 

\subsection{Residual WFE and contrast performance}
\label{sec:perfo}

Having determined the optimum integrator gains and split ratios for the CAO system, we can now compare the performance of both splitting concepts and quantify the correction improvement provided by the 2\nd stage. For the gray beam-splitting case, 80\% of the flux is sent to the 1\st stage and 20\% is sent to the 2\nd stage. This ratio is optimum for all but the faintest stars as shown in Sec.~\ref{sec:bs_gain}, and it is kept fixed because a single optical beam-splitters would not allow one to change the ratio depending on observing conditions. For the dichroic beam-splitting case, the star is assumed to have a red I-J color of 2.06, as motivated in Sec.~\ref{sec:bs_gain}. 

\begin{figure}[ht]
   \begin{center}
   \begin{tabular}{c}
   \includegraphics[height=6.9cm]{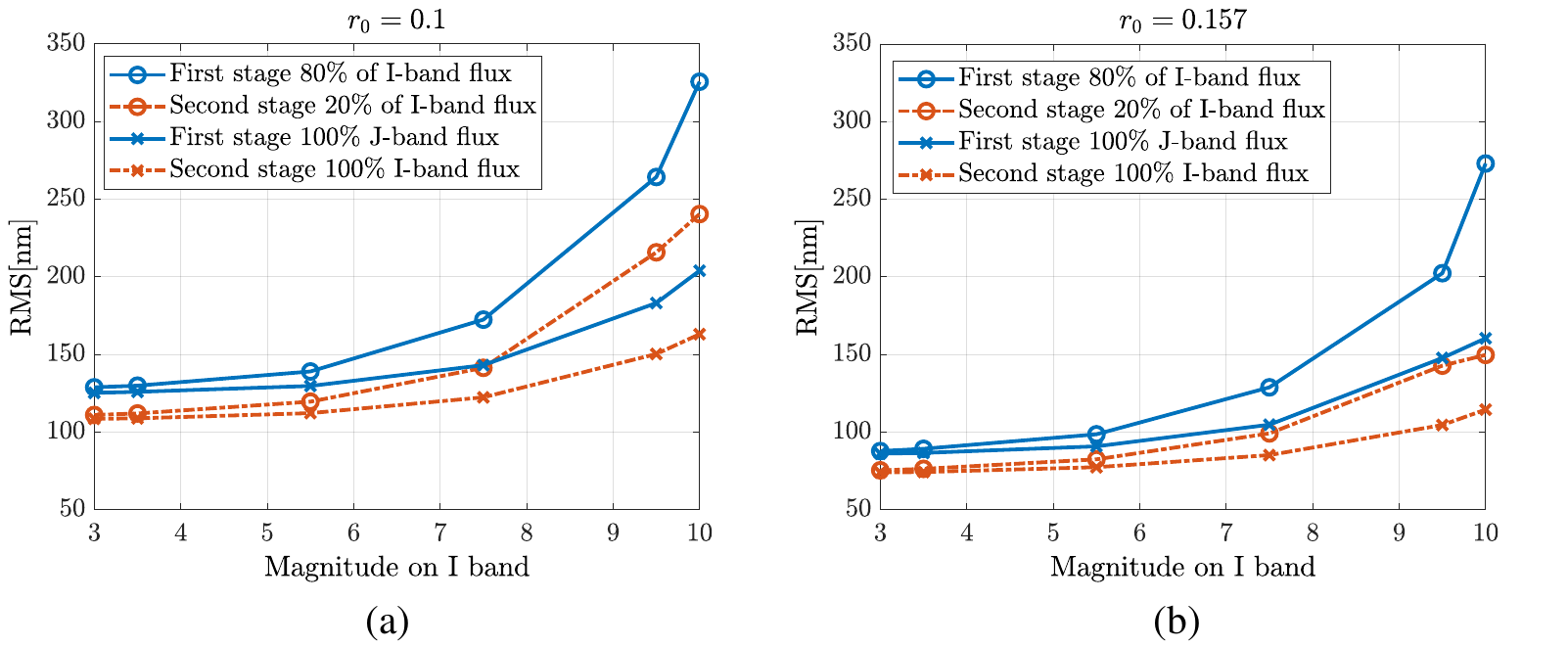}
   \end{tabular}
   \end{center}
   \caption [RMS in nanometers in terms of magnitude for the 1\st stage only (solid blue line) and with the 2\nd stage (dashed orange line) for (a) $r_0=0.1$ and (b) $r_0=0.157$. In the case where both wavefront sensors work on the I-band, the full flux is divided in a $80\%/20\%$ split ratio between both stages (circle markers). In the other case, the flux is divided in different wavelengths for each stage, and 100\% of the flux is sent (cross markers).]
   {\label{fig:RMS_STG1_STG2} RMS in nanometers in terms of magnitude for the 1\st stage only (solid blue line) and with the 2\nd stage (dashed orange line) for (a) $r_0=0.1$ and (b) $r_0=0.157$. In the case where both wavefront sensors work on the I-band, the full flux is divided in a $80\%/20\%$ split ratio between both stages (circle markers). In the other case, the flux is divided in different wavelengths for each stage, and 100\% of the flux is sent (cross markers).}
\end{figure}

Figure~\ref{fig:RMS_STG1_STG2} shows the CAO systems' residual WFE after 10,000 iterations on the 1\st stage (i.e., after 10 seconds of closed-loop operation) as a function of guide star I-band magnitude for good and median seeing values. Figure~\ref{fig:RMS_STG1_STG2} shows that the CAO (orange dashed line) consistently outperforms the single stage AO for all guide star magnitudes independently from how the beam-splitting is done. Not surprisingly, dichroic beam-splitting leads to a better correction performance overall because it provides more photons for each WFS individually, leading to a better correction performance overall. The curves exhibit the expected behavior of a rather constant bright guide star performance dominated by fitting error residuals and a degradation for fainter stars where measurement noise is dominating the error budget.

Besides the residual RMS WFE, the residual point spread function (PSF) contrast presents another important performance metric for high-contrast imaging. We calculate the residual PSF from the residual WFE assuming that Airy diffraction pattern has been removed by an idealized perfect coronagraph \cite{cavarroc_fundamental_2006}. The one-dimensional residual PSF contrast is then given by the standard deviation of the flux intensities in a thin annulus of a given angular radius normalized by the peak intensity of the non-coronagraphic PSF.

\begin{figure}[ht]
   \begin{center}
   \begin{tabular}{c} 
   \includegraphics[height=10cm]{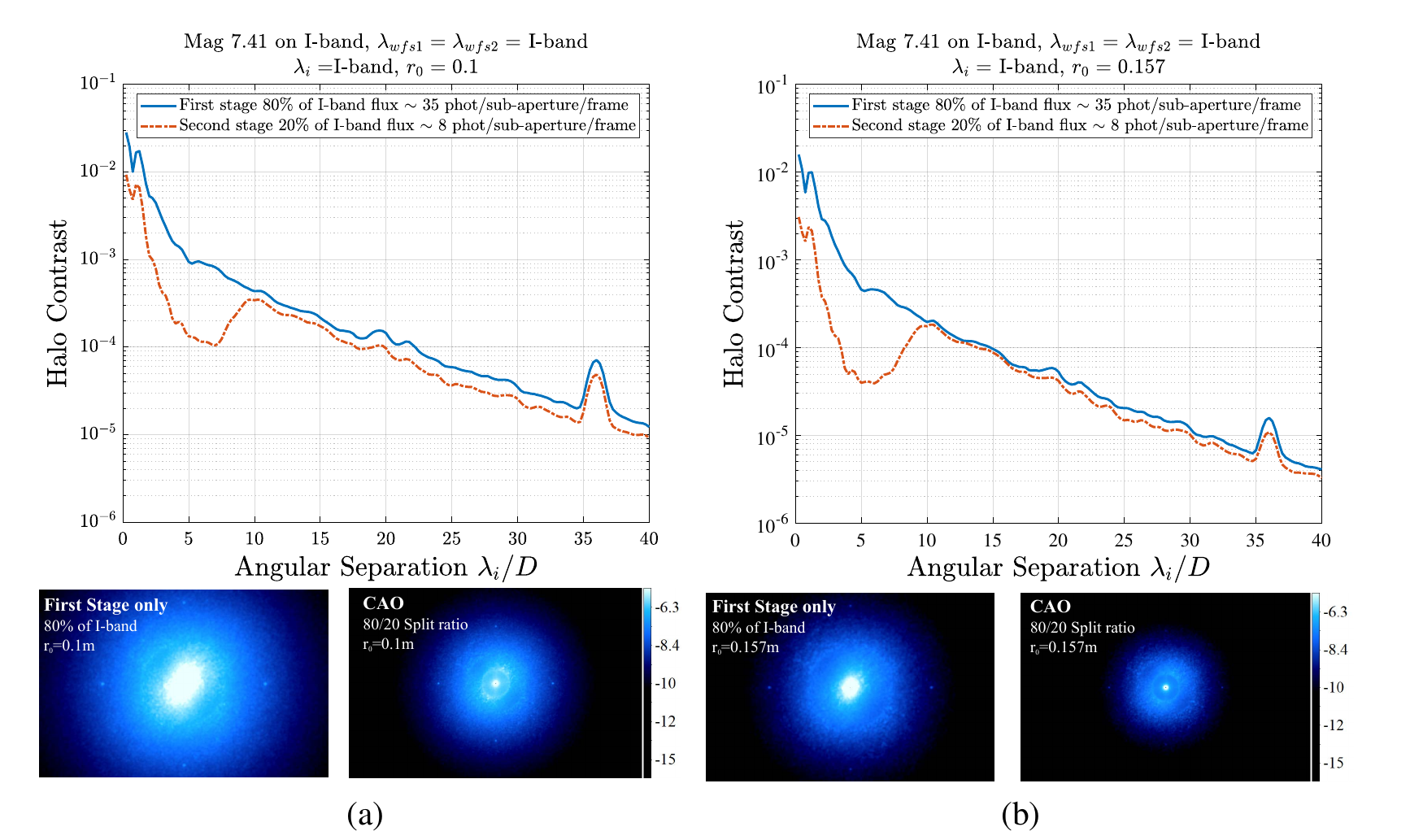} 
   \end{tabular}
   \end{center}
   \caption [Contrast performance and PSF for $r_0=0.1$ (a) and $r_0=0.157$ (b). Both stages operate their WFS in I-band, and the optimum gray beamsplitting ratio (80\%/20\%) was used. Scientific analysis is done in I-band as well.]
   { \label{fig:contrast_and_psf_split} Contrast performance and PSF for $r_0=0.1$ (a) and $r_0=0.157$ (b). Both stages operate their WFS in I-band, and the optimum gray beamsplitting ratio (80\%/20\%) was used. Scientific analysis is done in I-band as well.}
\end{figure}

\begin{figure}[ht]
   \begin{center}
   \begin{tabular}{c} 
   \includegraphics[height=13cm]{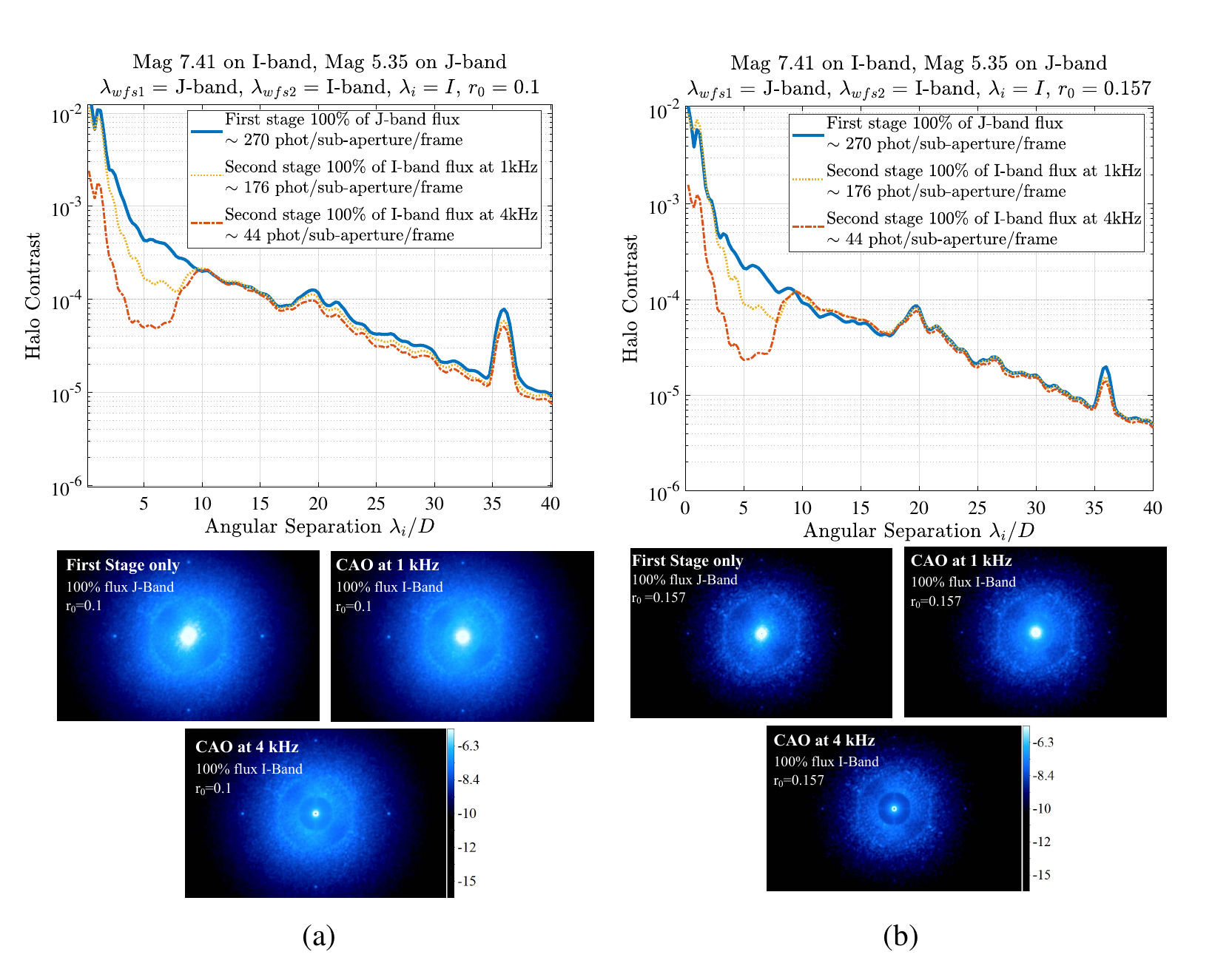}
   \end{tabular}
   \end{center}
   \caption[Contrast performance and PSF for $r_0=0.1$ (a) and $r_0=0.157$ (b) and dichroic beamsplitting. All the J-Band flux was sent to the 1\st stage, and all the I-band flux was sent to the 2\nd stage. Scientific analysis is done in I-band.] 
   { \label{fig:contrast_and_psf_proxb} Contrast performance and PSF for $r_0=0.1$ (a) and $r_0=0.157$ (b) and dichroic beamsplitting. All the J-Band flux was sent to the 1\st stage, and all the I-band flux was sent to the 2\nd stage. Scientific analysis is done in I-band.}
\end{figure}

Figure~\ref{fig:contrast_and_psf_split} shows the residual coronagraphic PSF and its contrast for the gray beam-splitting case in median and good seeing, and Fig.~\ref{fig:contrast_and_psf_proxb} shows the same for the dichroic beam-splitting case. The simulated observations represent the flux case of Proxima Centauri with I = 7.4, so neither very bright nor very faint. Imaging is done in I-band where the A-band of molecular oxygen is located. The improvement provided by the fast second correction stage is demonstrated by the better contrast in the controlled region of the 2\nd stage DM at separations smaller than about 9 $\lambda/D$. The 2\nd stage removes the elongated wind-driven halo \cite{cantalloube_f_wind-driven_2020} which is a signpost for the temporal error of the AO and improves the residual halo contrast at these angular separations by almost one order of magnitude. This is consistent with the analytical prediction that the temporal error is proportional to $f_c^{-5/3}$ (with $f_c$ denoting the correction frequency), so one would expect an about ten times improvement for a four times faster correction.

The best performance for the Proxima Centauri case is achieved by beam-splitting with a dichroic and reaches a contrast of $2\times10^{-4}$ for good seeing at the maximum angular separations of Proxima b of about 40 mas or 2 $\lambda/D$ for an 8-m telescope observing at 760 nm. Such a contrast performance should allow us to detect Oxygen in a hypothetical Proxima b atmosphere with an Earth-like composition in a few hundred hours \cite{lovis_atmospheric_2017}.Such an observation would require the AO WFS to work at a waveband slightly longer than I-band (e.g. between ~800nm and 950nm) to send all the 760 nm photons to the spectrograph..

Using a PWS for the fast second stage combines several beneficial effects at the same time. It reduces temporal error, aliasing error \cite{verinaud05} and noise error \cite{guyon_limits_2005}. The aliasing error of the 1\st stage SHS \cite{rigaut98} should be around 30nm and 45nm RMS for an $r_0$ of 0.157m and 0.1m, respectively, while the temporal errors \cite{fried90} for an assumed time delay of 2 frames, so 2ms, should be 50nm and  75nm. The photon-noise limited SHS centroiding error \cite{hardy98} assuming a noise propagation on $n$ reconstructed modes proportional to $log(n)$ \cite{rigaut98} yields about 50nm RMS. Therefore, aliasing and photon noise cannot be neglected and a significant contrast improvement is already expected when adding a slow 2\nd stage with a PWS. 

The yellow dotted line on Figure.~\ref{fig:contrast_and_psf_proxb} shows the simulated contrast from a “slow” 1kHz 2\nd stage PWS. The reduction of aliasing and the higher sensitivity of the PWS compared to the SHS already results in a significant gain in contrast of a factor 2-3. Another factor 3-5 is then gained by running the second stage faster and reduce the temporal error as seen by the red dotted line on the same figure.

\begin{figure}[ht]
   \begin{center}
   \begin{tabular}{c}
   \includegraphics[height=6cm]{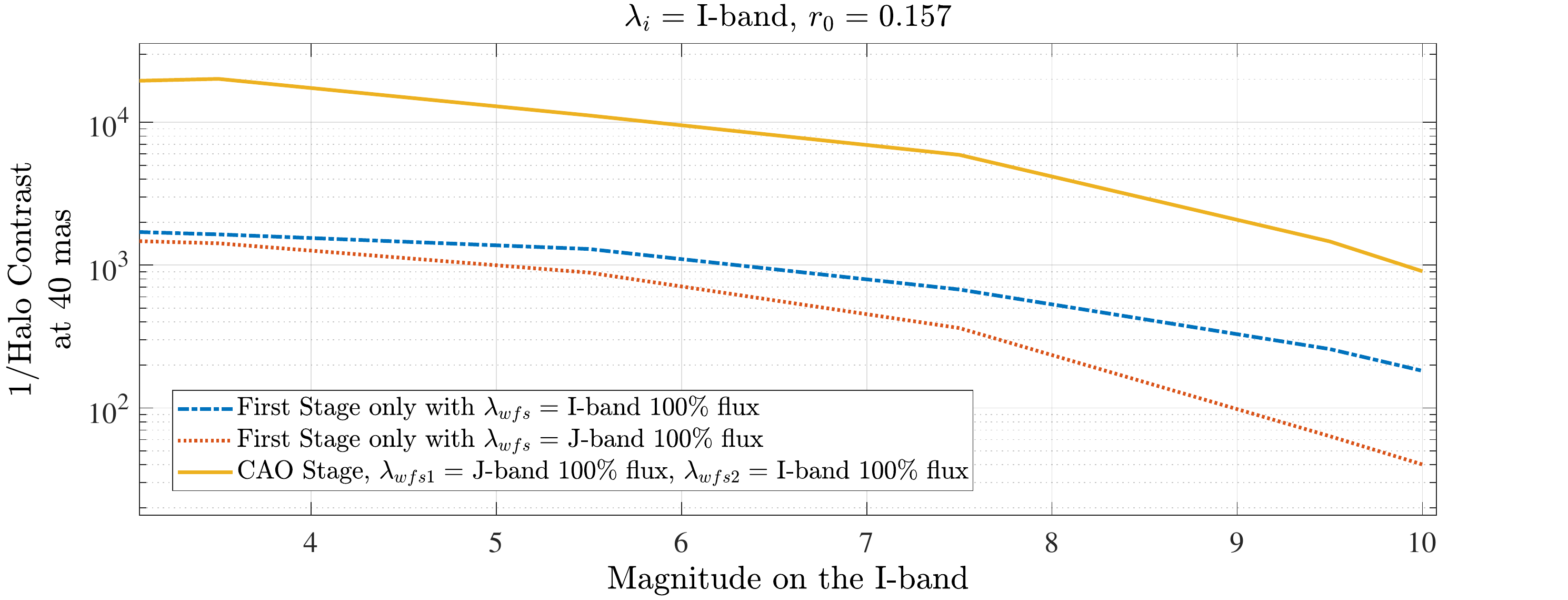} 
   \end{tabular}
   \end{center}
   \caption[PSF residual contrast as a function of magnitude at an angular separations of 40 mas.] 
   { \label{fig:contrast_mag} PSF residual contrast as a function of magnitude at an angular separations of 40 mas.}
\end{figure}

Figure.~\ref{fig:contrast_mag} shows how the contrast at 40 mas is improved by the 2\nd stage as a function of stellar I-band magnitude. Again, the CAO system provides a contrast improvement of roughly one order of magnitude when compared to the single stage AO. Similar to the residual wavefront error shown in Fig.~\ref{fig:RMS_STG1_STG2}, we see that the correction and contrast performance degrade with stellar magnitude due to the increased noise and reduced optimum integrator gains. In contrast to the residual wavefront error, the contrast in the bright end does not level out because of a dominating fitting error which occurs at spatial frequencies beyond the correction radius of the AO and would not affect the contrast at small angular separations. We rather see that aliasing and residual temporal error for the system approaching the maximum stable gain and thereby operating at its maximum correction bandwidth set the contrast cap for very bright stars. In the faint end, the CAO system degrades less rapidly because its 2\nd stage stage is operated by the more sensitive Pyramid WFS.

\subsection{Analysis of AO residual speckle lifetime}
\label{sec:speckle}

Another important parameter for HCI is the lifetime of speckles in the AO residual PSF. If we assume that the speckle noise would reduce with $1/\sqrt{t_{dc}}$, where $t_{dc}$ is the speckle decorrelation time, a residual speckle halo with typical intensity contrast of about $10^{-5}$ would require $10^6$ independent realizations to reach a level of $10^{-8}$ in the absence of other speckle correction techniques such as angular, spectral or polarimetric differential imaging. Depending on the speckle lifetime, accumulating that many realizations of the speckle pattern can be a very long process. 

Long-lived quasi-static speckles are produced by instrument aberrations, which are left uncorrected by the AO system \cite{marois_effects_2003} and are one of the dominating factors affecting contrast especially at low angular separations \cite{guyon_limits_2005}. These residual aberrations can occur in the instrument's science camera optical path, which is not seen by the AO WFS, or in the optical path to the AO WFS, which is not seen by the science camera. Therefore they are called non-common path aberrations (NCPA). NCPA change only slowly on timescales on which the instrument orientation and the gravity vector changes during an observation tracking a target in the sky.  Also, temperature variations that produce thermal expansions in the instrument may introduce NCPA. Residuals from the atmospheric turbulence can induce a fast partial decorrelation of the PSF over a few seconds before transiting to a linear decorrelation regime at small angular separations\cite{milli_adaptive_2016}. A refined analysis further revealed another speckle decorrelation time scale of less than 2 ms which can be attributed to the AO correction\cite{goebel_measurements_2018}.

Our simulations do not include NCPA, so we are solely looking at the temporal evolution of residual atmospheric turbulence speckle intensities. Assuming Taylor's frozen flow hypothesis, it was shown in Ref.~\citenum{macintosh_speckle_2005} that an integrator-controlled AO system does not change the speckle lifetime compared to uncorrected turbulence but leads to an overall reduction in speckle intensity. This is explained by the idea that the correction always trails behind incoming turbulence leaving a residual with unchanged temporal characteristics. Moreover, the speckle lifetime is proportional to the ratio between telescope diameter ($D$) and wind-speed ($v$), more specifically $0.6 D/v$. In the case of the VLT, with winds of $10$ m/s, the atmospheric residual speckle lifetime is therefore of the order of half a second, while speckles are expected to decorrelate on timescales of several seconds in the ELT case. Such lifetimes would lead to unfeasible long exposure time requirements (more than 100 hours for the $10^6$ independent realizations motivated above) for reaching very high contrast. These considerations underline the high interest in reducing the lifetime of residual atmospheric speckles, even if the frozen flow assumption may be pessimistic in this context. 

\begin{figure}[ht]
   \begin{center}
   \begin{tabular}{c} 
   \includegraphics[height=5cm]{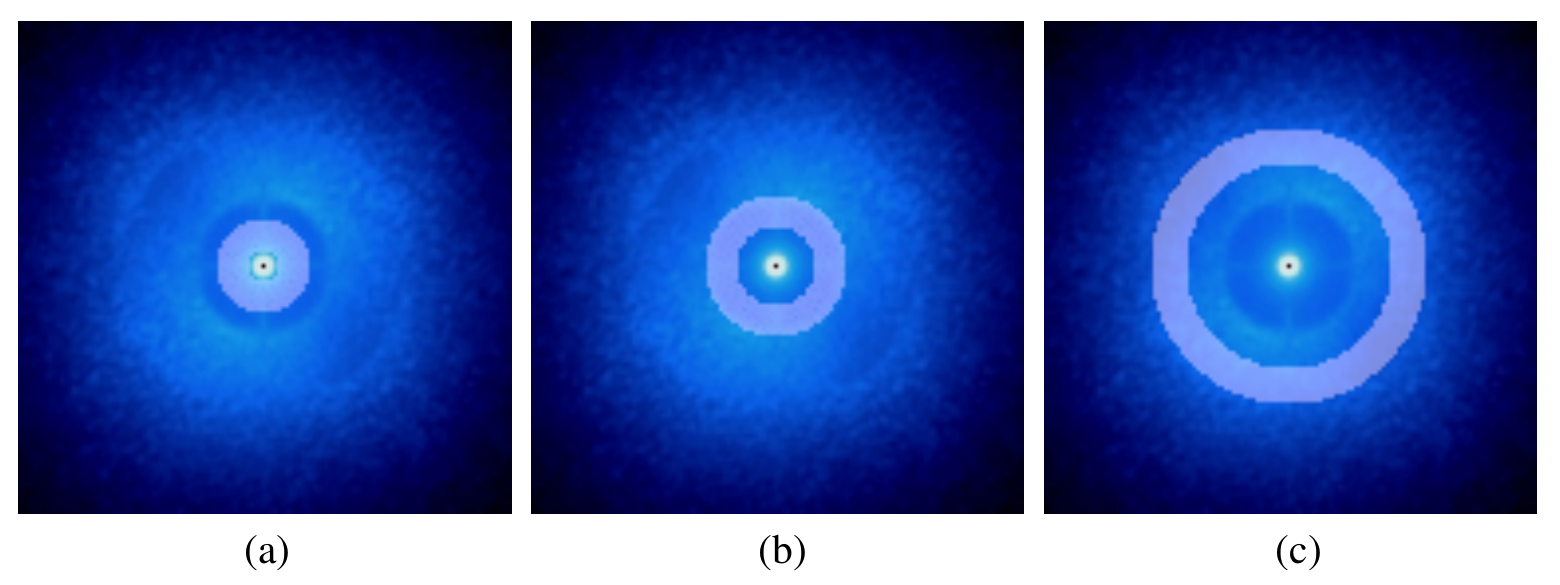} \\
   \end{tabular}
   \end{center}
   \caption[Long exposure perfect coronagraph PSF with different regions $A$ over imposed] 
   { \label{fig:maskPSF}Long exposure perfect coronagraph PSF with different regions $A$ over imposed. (a):~$A_1= 2\textup{--}5[\frac{\lambda}{D}]$ (b):~$A_2=  5\textup{--}8[\frac{\lambda}{D}]$ and (c):~$A_3= 12\textup{--}15[\frac{\lambda}{D}]$.}
\end{figure}

In Sec.~\ref{sec:control}, we showed that the low frequency part of the CAO's loop correction transfer function is very similar to the one of a double-integrator controller. The CAO system therefore presents a much more efficient reduction of the aberration energy at low temporal frequencies, and an effect on the speckle lifetime should be observable in the image plane. Therefore, we apply published analysis \cite{milli_speckle_2016} on our simulated coronagraphic images and compare speckle lifetimes for single and double stage AO correction. We analysed three annular regions at different angular separations from the PSF center: $A_1= 2\textup{--}5[\frac{\lambda}{D}]$, $A_2= 5\textup{--}8[\frac{\lambda}{D}]$ and $A_3= 12\textup{--}15[\frac{\lambda}{D}]$ as indicated in Fig.~\ref{fig:maskPSF}. While $A_1$ and $A_2$ are inside the correction radius and controlled by both stages of the CAO, $A_3$ is only affected by the 1\st stage and could therefore show a different speckle lifetime. We simulated a short 2.5 seconds observation with $r_0$ = 0.157m and an elevated wind speed, $50\%$ higher than the wind speed used for the performance simulations (See Table~\ref{tab:System-Parameters}).  For each of the three regions, we arranged the 2.5 seconds worth of imaging data in a matrix which contains the evolution over time for each pixel. We then subtracted the mean intensity of each pixel and calculated the temporal autocorrelation functions. Finally, the autocorrelation functions of all the pixels were averaged to derive the typical temporal correlation of the residual speckles in the three considered regions.

\begin{figure}[ht]
   \begin{center}
   \begin{tabular}{c}
   \includegraphics[height=11cm]{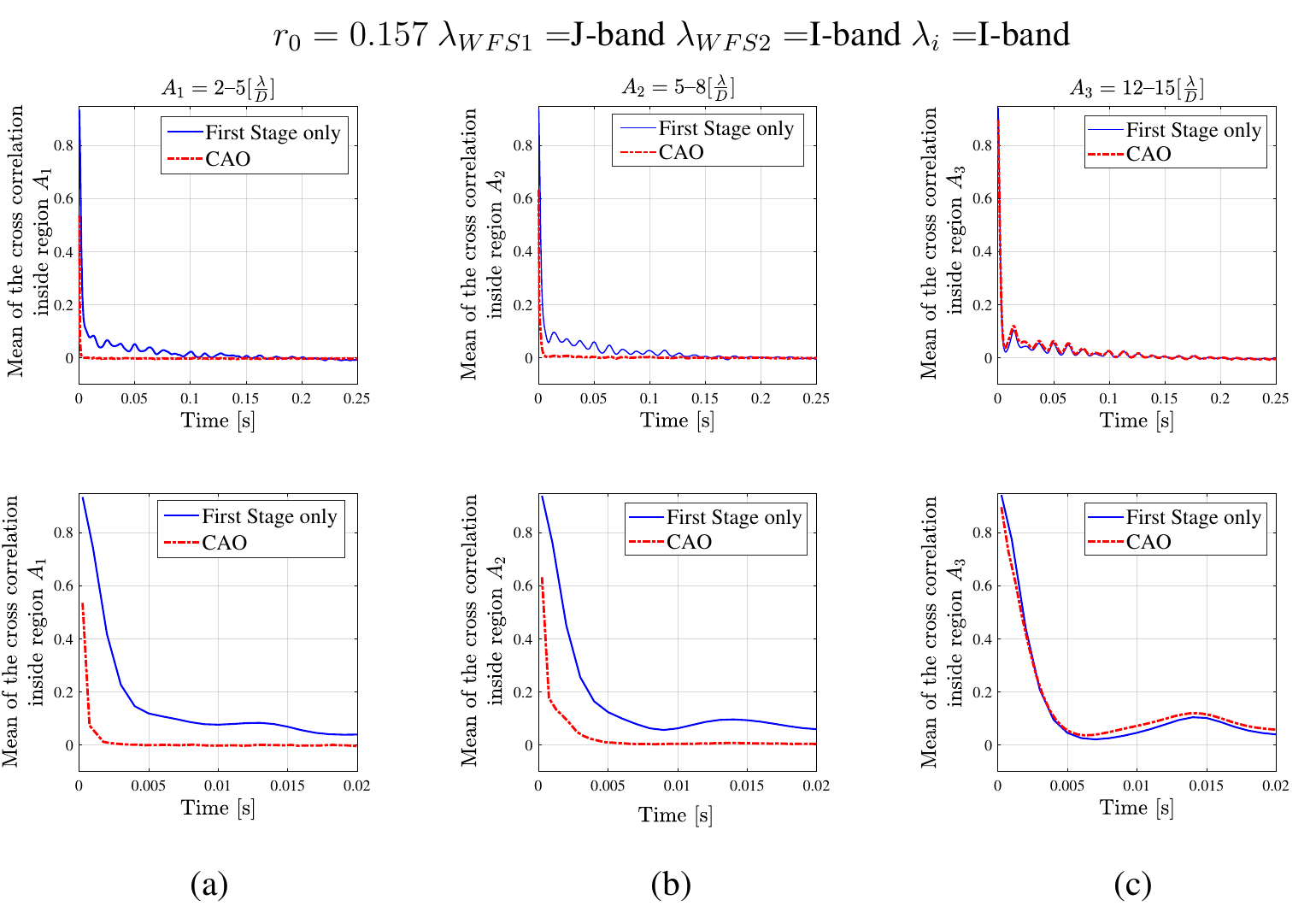} \\
   \end{tabular}
   \end{center}
   \caption[Temporal decorrelations of all the pixels inside $A$ for three different regions] 
   { \label{fig:CORRCOEFF}Temporal decorrelations of all the pixels inside $A$ for three different regions: (a) $A_1= 2\textup{--}5[\frac{\lambda}{D}]$, (b) $A_2= 5\textup{--}8[\frac{\lambda}{D}]$ and (c) $A_3= 12\textup{--}15[\frac{\lambda}{D}]$. The panels on the bottom zoom in on very short timescales.}
\end{figure}

The results are shown Fig.~\ref{fig:CORRCOEFF}. In the regions $A_1$ and $A_2$ we clearly see the effect of the 2\nd stage. While the lifetime of residual speckles after the 1\st stage is of the order of $0.15$ seconds, it is reduced to just a few ms after the 2\nd stage. This reduction of a factor 30-50 is much larger than the fourfold increased correction speed offered by the 2\nd stage. We also see that the decorrelation time in region $A_3$ is not affected by the CAO. This is the expected behavior because $A_3$ is outside of the control region of the 2\nd stage. Compared to the relatively long lifetime of residual atmospheric speckle of a single stage AO, the fast decorrelation of residual speckles of the CAO will help to smoothen the residual PSF efficiently. The more than 100 hours for $10^6$ independent realizations would then shorten to just a few hours, which is consistent with typical HCI observing times.

\section{Summary and discussion}
\label{sec:concl}

A two-stage CAO system with a faster 2nd stage presents an efficient way to significantly improve the contrast performance of an existing XAO system limited by temporal error which is of paramount interest for the Exoplanet science case. While a single stage XAO system running at high speed and equipped with a high order DM could probably provide better results than the proposed CAO system, it would require a more powerful RTC and a fast response and large stroke of the high-order DM. The proposed CAO system instead provides a cost and resource effective solution to improve the performance of an existing XAO system. Further benefits are that the 2\nd stage can easily be integrated and tested stand-alone and retrofitted into an existing instrument. The main complication introduced by this concept is that the 1\st stage increases the relative content of high temporal frequency disturbance input to the 2\nd stage because of its controller overshoot and the inter-sampling signal (see Sec.\ref{sec:control}). Possibilities to mitigate this effect include running the 1\st stage at a reduced gain in noisy conditions, or designing the 2\nd stage controller in a way that effectively copes with the inter-sampling signal. A straightforward solution would be to include a notch filter in the 2\nd stage loop. Also, a global design of an optimized 2\nd stage predictive controller would be of great interest. These developments deserve a complete study and are left for future work. We also showed that the low temporal frequency rejection of the CAO is partly the same than that of a SCAO system controlled with a double-integrator and is much higher than that of a SCAO system controlled with a standard integrator.

Numerical simulations of a 1st stage SHS AO system running at 1 kHz (similar to the existing VLT-SPHERE or AOF) show that a fast 2\nd stage employing a sensitive WFS such as the unmodulated PWS would improve the correction performance for all GS magnitudes. A simple integrator control for both stages where the 2\nd stage runs at four times the framerate of the first one increases the contrast by about one order of magnitude which would translate into a similar reduction of science exposure time required to reach a certain S/N. The integrator gains of both stages must be jointly optimized to reach optimum performance to cope with the 1\st stage altering the input disturbance to the 2\nd stage. The two stage CAO can provide at 2 $\lambda/D$ an I-band contrast of the order 1:5000 for an 8-m telescope with a 1\st stage AO correcting 800 modes at a framerate of 1 kHz. Such a performance would for example bring the detection of oxygen in the atmosphere of Proxima b (if it were present at an Earth-like abundance) within reach.

AO systems which already use the sensitive PWS and run at high framerates such as KPIC \cite{mawet_keck_2018} or MagAO-X \cite{Males20} are already able to reduce all relevant AO residual error terms (aliasing, temporal, and noise errors) to the minimum and therefore wouldn't benefit from a CAO architecture in the same way as the system we have studied. In particular, the double-integrator behavior of the CAO for low frequencies can also be obtained by a single-stage AO.

Finally, a CAO with a four times faster 2\nd stage reduces the decorrelation or lifetime of atmospheric turbulence speckles by a factor 30-50 over the lifetime observed with the 1\st stage only. While this obviously does not reduce photon noise, it helps to smoothen the residual halo more rapidly and reduce its ``granularity". This CAO therefore leads to a reduction of atmospheric speckle noise such that spatial low-pass filtering methods can be used to improve the final image contrast effectively.

\subsection* {Acknowledgments}
This work has received funding from the European Unions Horizon 2020 research and innovation program, grant agreement No 730890.

%%%%% References %%%%%

\bibliography{report}   % bibliography data in report.bib

\begin{thebibliography}{10}

\bibitem{mayor_jupiter-mass_1995}
M.~Mayor and D.~Queloz, ``A {Jupiter}-mass companion to a solar-type star,''
  {\em {Nature}} {\bf 378}, 355--359  (1995).

\bibitem{marois_direct_2008}
C.~Marois, B.~Macintosh, T.~Barman, {\em et~al.}, ``Direct {Imaging} of
  {Multiple} {Planets} {Orbiting} the {Star} {HR} 8799,'' {\em {Science}} {\bf
  322}, 1348  (2008).

\bibitem{marois_images_2010}
C.~Marois, B.~Zuckerman, Q.~M. Konopacky, {\em et~al.}, ``Images of a fourth
  planet orbiting {HR} 8799,'' {\em {Nature}} {\bf 468}, 1080--1083  (2010).

\bibitem{lagrange_giant_2010}
A.-M. Lagrange, M.~Bonnefoy, G.~Chauvin, {\em et~al.}, ``A {Giant} {Planet}
  {Imaged} in the {Disk} of the {Young} {Star} beta {Pictoris},'' {\em Science}
  {\bf 329}, 57  (2010).

\bibitem{rameau_discovery_2013}
J.~Rameau, G.~Chauvin, A.-M. Lagrange, {\em et~al.}, ``Discovery of a
  {Probable} 4-5 {Jupiter}-mass {Exoplanet} to {HD} 95086 by {Direct}
  {Imaging},'' {\em {ApJ}l} {\bf 772}, L15  (2013).

\bibitem{bailey_hd_2014}
V.~Bailey, T.~Meshkat, M.~Reiter, {\em et~al.}, ``{HD} 106906 b: {A}
  {Planetary}-mass {Companion} {Outside} a {Massive} {Debris} {Disk},'' {\em
  {ApJ}l} {\bf 780}, L4  (2014).

\bibitem{macintosh_discovery_2015}
B.~Macintosh, J.~R. Graham, T.~Barman, {\em et~al.}, ``Discovery and
  spectroscopy of the young jovian planet 51 {Eri} b with the {Gemini} {Planet}
  {Imager},'' {\em Science} {\bf 350}, 64--67  (2015).

\bibitem{chauvin_discovery_2017}
G.~Chauvin, S.~Desidera, A.-M. Lagrange, {\em et~al.}, ``Discovery of a warm,
  dusty giant planet around {HIP} 65426,'' {\em A\&A} {\bf 605}, L9  (2017).
\newblock Publisher: EDP Sciences.

\bibitem{otten_direct_2021}
G.~P. P.~L. Otten, A.~Vigan, E.~Muslimov, {\em et~al.}, ``Direct
  characterization of young giant exoplanets at high spectral resolution by
  coupling {SPHERE} and {CRIRES}+,'' {\em {A\&A}} {\bf 646}, A150  (2021).
\newblock \_eprint: 2009.01841.

\bibitem{guyon_limits_2005}
O.~Guyon, ``Limits of {Adaptive} {Optics} for {High}‐{Contrast} {Imaging},''
  {\em {ApJ}} {\bf 629}, 592--614  (2005).

\bibitem{chilcote_gpi_2020}
J.~Chilcote, Q.~Konopacky, R.~J.~D. Rosa, {\em et~al.}, ``{GPI} 2.0: upgrading
  the {Gemini} {Planet} {Imager},'' in {\em Ground-based and {Airborne}
  {Instrumentation} for {Astronomy} {VIII}},  C.~J. Evans, J.~J. Bryant, and
  K.~Motohara, Eds.,  {\bf 11447}, 394 -- 407, SPIE  (2020).

\bibitem{males_ground-based_2018}
J.~R. Males and O.~Guyon, ``Ground-based adaptive optics coronagraphic
  performance under closed-loop predictive control,'' {\em Journal of
  Astronomical Telescopes, Instruments, and Systems} {\bf 4}, 019001  (2018).

\bibitem{bond_adaptive_2020}
C.~Z. Bond, S.~Cetre, S.~Lilley, {\em et~al.}, ``Adaptive optics with an
  infrared pyramid wavefront sensor at {Keck},'' {\em {JATIS}} {\bf 6}(3), 1 --
  21  (2020).

\bibitem{lozi_scexao_2018}
J.~Lozi, O.~Guyon, N.~Jovanovic, {\em et~al.}, ``{SCExAO}, an instrument with a
  dual purpose: perform cutting-edge science and develop new technologies,'' in
  {\em Adaptive Optics Systems {VI}},  L.~M. Close, L.~Schreiber, and
  D.~Schmidt, Eds., {\em Society of Photo-Optical Instrumentation Engineers
  ({SPIE}) Conference Series} {\bf 10703}, 1070359  (2018).
\newblock \_eprint: 1809.08301.

\bibitem{boccaletti_sphere_2020}
A.~Boccaletti, G.~Chauvin, D.~Mouillet, {\em et~al.}, ``{SPHERE+: Imaging young
  Jupiters down to the snowline}.'' arXiv: Earth and Planetary Astrophysics
  (2020).

\bibitem{chazelas_ristretto_2020-1}
B.~Chazelas, C.~Lovis, N.~Blind, {\em et~al.}, ``{RISTRETTO}: a pathfinder
  instrument for exoplanet atmosphere characterization,'' in {\em Adaptive
  {Optics} {Systems} {VII}},   {\bf 11448}, 1144875, International Society for
  Optics and Photonics  (2020).

\bibitem{perrin_structure_2003}
M.~D. Perrin, A.~Sivaramakrishnan, R.~B. Makidon, {\em et~al.}, ``The
  {Structure} of {High} {Strehl} {Ratio} {Point}-{Spread} {Functions},'' {\em
  {ApJ}} {\bf 596}, 702--712  (2003).

\bibitem{beuzit_sphere_2019-1}
J.~L. Beuzit, A.~Vigan, D.~Mouillet, {\em et~al.}, ``{SPHERE}: the exoplanet
  imager for the {Very} {Large} {Telescope},'' {\em {A\&A}} {\bf 631}, A155
  (2019).
\newblock \_eprint: 1902.04080.

\bibitem{madec_adaptive_2018}
P.~Y. Madec, R.~Arsenault, H.~Kuntschner, {\em et~al.}, ``Adaptive {Optics}
  {Facility}: from an amazing present to a brilliant future...,'' in {\em
  Adaptive {Optics} {Systems} {VI}},  L.~M. Close, L.~Schreiber, and
  D.~Schmidt, Eds., {\em Society of {Photo}-{Optical} {Instrumentation}
  {Engineers} ({SPIE}) {Conference} {Series}} {\bf 10703}, 1070302  (2018).

\bibitem{dessenne_modal_1997}
C.~Dessenne, P.-Y. Madec, and G.~Rousset, ``Modal prediction for closed-loop
  adaptive optics,'' {\em Optics Letters} {\bf 22}, 1535--1537  (1997).

\bibitem{gavel_toward_2003}
D.~T. Gavel and D.~Wiberg, ``Toward {Strehl}-optimizing adaptive optics
  controllers,'' in {\em Adaptive {Optical} {System} {Technologies} {II}},
  P.~L. Wizinowich and D.~Bonaccini, Eds., {\em Society of {Photo}-{Optical}
  {Instrumentation} {Engineers} ({SPIE}) {Conference} {Series}} {\bf 4839},
  890--901  (2003).

\bibitem{le_roux_optimal_2004}
B.~Le~Roux, J.-M. Conan, C.~Kulcs{\'a}r, {\em et~al.}, ``Optimal control law
  for classical and multiconjugate adaptive optics,'' {\em Journal of the
  Optical Society of America A} {\bf 21}, 1261  (2004).

\bibitem{poyneer_fourier_2007}
L.~A. Poyneer, B.~A. Macintosh, and J.-P. Véran, ``Fourier transform wavefront
  control with adaptive prediction of the atmosphere,'' {\em Journal of the
  Optical Society of America A} {\bf 24}, 2645  (2007).

\bibitem{piatrou_performance_2007}
P.~Piatrou and M.~C. Roggemann, ``Performance study of {Kalman} filter
  controller for multiconjugate adaptive optics,'' {\em Applied optics} {\bf
  46}(9), 1446--1455  (2007).
\newblock Publisher: Optical Society of America.

\bibitem{hinnen_exploiting_2007}
K.~Hinnen, M.~Verhaegen, and N.~Doelman, ``Exploiting the spatiotemporal
  correlation in adaptive optics using data-driven {H}$_2$-optimal control,''
  {\em JOSA A} {\bf 24}(6), 1714--1725  (2007).
\newblock Publisher: Optical Society of America.

\bibitem{petit_linear_2009}
C.~Petit, J.-M. Conan, C.~Kulcsár, {\em et~al.}, ``Linear quadratic {Gaussian}
  control for adaptive optics and multiconjugate adaptive optics: experimental
  and numerical analysis,'' {\em J. Opt. Soc. Am. A} {\bf 26}, 1307--1325
  (2009).
\newblock Publisher: OSA.

\bibitem{fraanje_fast_2010}
R.~Fraanje, J.~Rice, M.~Verhaegen, {\em et~al.}, ``Fast reconstruction and
  prediction of frozen flow turbulence based on structured {Kalman}
  filtering,'' {\em JOSA A} {\bf 27}, A235--A245  (2010).
\newblock Publisher: OSA.

\bibitem{kulcsar_minimum_2012}
C.~Kulcsár, H.-F. Raynaud, C.~Petit, {\em et~al.}, ``Minimum variance
  prediction and control for adaptive optics,'' {\em Automatica} {\bf 48},
  1939--1954  (2012).

\bibitem{correia_spatio-angular_2015}
C.~M. Correia, K.~Jackson, J.-P. Véran, {\em et~al.}, ``Spatio-angular
  minimum-variance tomographic controller for multi-object adaptive-optics
  systems,'' {\em {Appl.~Opt.}} {\bf 54}, 5281  (2015).

\bibitem{guyon_adaptive_2017}
O.~Guyon and J.~Males, ``Adaptive optics predictive control with empirical
  orthogonal functions (eofs).'' arXiv: Instrumentation and Methods for
  Astrophysics  (2017).

\bibitem{gluck_model_2018}
M.~Glück, J.~Pott, and O.~Sawodny, ``Model {Predictive} {Control} of
  {Multi}-{Mirror} {Adaptive} {Optics} {Systems},'' in {\em 2018 {IEEE}
  {Conference} on {Control} {Technology} and {Applications} ({CCTA})},
  909--914  (2018).

\bibitem{prengere_zonal-based_2020}
L.~Prengère, C.~Kulcsár, and H.-F. Raynaud, ``Zonal-based high-performance
  control in adaptive optics systems with application to astronomy and
  satellite tracking,'' {\em JOSA A} {\bf 37}(7), 1083--1099  (2020).
\newblock Publisher: Optical Society of America.

\bibitem{liu_wavefront_2020}
X.~Liu, T.~Morris, C.~Saunter, {\em et~al.}, ``Wavefront prediction using
  artificial neural networks for open-loop adaptive optics,'' {\em Monthly
  Notices of the Royal Astronomical Society} {\bf 496}, 456--464  (2020).

\bibitem{dessenne_sky_1999}
C.~Dessenne, P.-Y. Madec, and G.~Rousset, ``Sky implementation of modal
  predictive control in adaptive optics,'' {\em Optics letters} {\bf 24}(5),
  339--341  (1999).
\newblock Publisher: Optical Society of America.

\bibitem{doelman_real-sky_2011}
N.~Doelman, R.~Fraanje, and R.~den Breeje, ``Real-sky adaptive optics
  experiments on optimal control of tip-tilt modes,'' in {\em Second
  {International} {Conference} on {Adaptive} {Optics} for {Extremely} {Large}
  {Telescopes}.},  51  (2011).

\bibitem{sivo_first_2014}
G.~Sivo, C.~Kulcsár, J.-M. Conan, {\em et~al.}, ``First on-sky {SCAO}
  validation of full {LQG} control with vibration mitigation on the {CANARY}
  pathfinder,'' {\em Optics express} {\bf 22}(19), 23565--23591  (2014).
\newblock Publisher: Optical Society of America.

\bibitem{lardiere_multi-object_2014}
O.~Lardière, D.~Andersen, C.~Blain, {\em et~al.}, ``Multi-object adaptive
  optics on-sky results with {Raven},'' in {\em Adaptive {Optics} {Systems}
  {IV}},   {\bf 9148}, 91481G, International Society for Optics and Photonics
  (2014).

\bibitem{sinquin_-sky_2020}
B.~Sinquin, L.~Prengere, C.~Kulcsár, {\em et~al.}, ``On-sky results for
  adaptive optics control with data-driven models on low-order modes,'' {\em
  {MNRAS}} {\bf 498}(3), 3228--3240  (2020).
\newblock Publisher: Oxford University Press.

\bibitem{petit_sphere_2014}
C.~Petit, J.-F. Sauvage, T.~Fusco, {\em et~al.}, ``{SPHERE} {eXtreme} {AO}
  control scheme: final performance assessment and on sky validation of the
  first auto-tuned {LQG} based operational system,'' in {\em Adaptive {Optics}
  {Systems} {IV}},   {\bf 9148}, 91480O, International Society for Optics and
  Photonics  (2014).

\bibitem{poyneer_performance_2016}
L.~A. Poyneer, D.~W. Palmer, B.~Macintosh, {\em et~al.}, ``Performance of the
  {Gemini} {Planet} {Imager}’s adaptive optics system,'' {\em Applied Optics}
  {\bf 55}(2), 323--340  (2016).
\newblock Publisher: Optical Society of America.

\bibitem{correia_modeling_2017}
C.~M. Correia, C.~Z. Bond, J.-F. Sauvage, {\em et~al.}, ``Modeling astronomical
  adaptive optics performance with temporally filtered {Wiener} reconstruction
  of slope data,'' {\em {JOSAA}} {\bf 34}, 1877  (2017).

\bibitem{males_ground-based_2018-1}
J.~R. Males and O.~Guyon, ``Ground-based adaptive optics coronagraphic
  performance under closed-loop predictive control,'' {\em {JATIS}} {\bf 4}, 1
  (2018).

\bibitem{ragazzoni_sensitivity_1999}
R.~Ragazzoni and J.~Farinato, ``Sensitivity of a pyramidic {Wave} {Front}
  sensor in closed loop {Adaptive} {Optics},'' {\em {A\&A}} {\bf 350}, L23--L26
   (1999).

\bibitem{juvenal_linear_2018}
R.~Juvénal, C.~Kulcsár, H.-F. Raynaud, {\em et~al.}, ``Linear controller
  error budget assessment for classical adaptive optics systems,'' {\em Journal
  of the Optical Society of America A} {\bf 35}, 1465  (2018).

\bibitem{Conan_1995}
J.~M. {Conan}, G.~{Rousset}, and P.~Y. {Madec}, ``{Wave-front temporal spectra
  in high-resolution imaging through turbulence.},'' {\em JOSAA} {\bf 12},
  1559--1570  (1995).

\bibitem{verinaud_nature_2004}
C.~Vérinaud, ``On the nature of the measurements provided by a pyramid
  wave-front sensor,'' {\em Optics Communications} {\bf 233}, 27--38  (2004).

\bibitem{fusco_final_2014}
T.~Fusco, J.-F. Sauvage, C.~Petit, {\em et~al.}, ``Final performance and
  lesson-learned of {SAXO}, the {VLT}-{SPHERE} extreme {AO}: from early design
  to on-sky results,'' in {\em Society of {Photo}-{Optical} {Instrumentation}
  {Engineers} ({SPIE}) {Conference} {Series}},  {\em Society of
  {Photo}-{Optical} {Instrumentation} {Engineers} ({SPIE}) {Conference}
  {Series}} {\bf 9148}, 1  (2014).

\bibitem{Madec_AOF_2018}
P.-Y. Madec, R.~Arsenault, H.~Kuntschner, {\em et~al.}, ``{Adaptive Optics
  Facility: from an amazing present to a brilliant future...},'' in {\em
  Adaptive Optics Systems VI},  L.~M. Close, L.~Schreiber, and D.~Schmidt,
  Eds.,  {\bf 10703}, 1 -- 13, International Society for Optics and Photonics,
  SPIE  (2018).

\bibitem{anglada-escude_terrestrial_2016}
G.~Anglada-Escudé, P.~J. Amado, J.~Barnes, {\em et~al.}, ``A terrestrial
  planet candidate in a temperate orbit around {Proxima} {Centauri},'' {\em
  {Nature}} {\bf 536}, 437--440  (2016).

\bibitem{ribas_candidate_2018}
I.~Ribas, M.~Tuomi, A.~Reiners, {\em et~al.}, ``A candidate super-{Earth}
  planet orbiting near the snow line of {Barnard}'s star,'' {\em Nature} {\bf
  563}, 365--368  (2018).

\bibitem{diaz_sophie_2019}
R.~F. Díaz, X.~Delfosse, M.~J. Hobson, {\em et~al.}, ``The {SOPHIE} search for
  northern extrasolar planets. {XIV}. {A} temperate ({T}$_{\textrm{eq}} \sim$
  300 {K}) super-earth around the nearby star {Gliese} 411,'' {\em {A\&A}} {\bf
  625}, A17  (2019).

\bibitem{zechmeister_carmenes_2019}
M.~Zechmeister, S.~Dreizler, I.~Ribas, {\em et~al.}, ``The {CARMENES} search
  for exoplanets around {M} dwarfs. {Two} temperate {Earth}-mass planet
  candidates around {Teegarden}'s {Star},'' {\em {A\&A}} {\bf 627}, A49
  (2019).

\bibitem{quirrenbach_carmenes_2018}
A.~Quirrenbach, P.~J. Amado, I.~Ribas, {\em et~al.}, ``{CARMENES}:
  high-resolution spectra and precise radial velocities in the red and
  infrared,'' in {\em Ground-based and {Airborne} {Instrumentation} for
  {Astronomy} {VII}},  C.~J. Evans, L.~Simard, and H.~Takami, Eds., {\em
  Society of {Photo}-{Optical} {Instrumentation} {Engineers} ({SPIE})
  {Conference} {Series}} {\bf 10702}, 107020W  (2018).

\bibitem{wildi_nirps_2017}
F.~Wildi, N.~Blind, V.~Reshetov, {\em et~al.}, ``{NIRPS}: an adaptive-optics
  assisted radial velocity spectrograph to chase exoplanets around {M}-stars,''
  in {\em Society of {Photo}-{Optical} {Instrumentation} {Engineers} ({SPIE})
  {Conference} {Series}},  {\em Society of {Photo}-{Optical} {Instrumentation}
  {Engineers} ({SPIE}) {Conference} {Series}} {\bf 10400}, 1040018  (2017).

\bibitem{conan_object-oriented_2014}
R.~Conan and C.~Correia, ``Object-oriented {Matlab} adaptive optics toolbox,''
  in {\em Adaptive optics systems {IV}},   {\bf 9148}, 91486C, International
  Society for Optics and Photonics  (2014).

\bibitem{gendron_astronomical_1994}
E.~Gendron and P.~Lena, ``Astronomical adaptive optics. {I}. {Modal} control
  optimization.,'' {\em {A\&A}} {\bf 291}, 337--347  (1994).

\bibitem{korkiakoski_applying_2008}
V.~Korkiakoski, C.~Vérinaud, and M.~Le~Louarn, ``Applying sensitivity
  compensation for pyramid wavefront sensor in different conditions,'' in {\em
  Adaptive {Optics} {Systems}},  N.~Hubin, C.~E. Max, and P.~L. Wizinowich,
  Eds., {\em Society of {Photo}-{Optical} {Instrumentation} {Engineers}
  ({SPIE}) {Conference} {Series}} {\bf 7015}, 701554  (2008).

\bibitem{deo_modal_2018}
V.~Deo, E.~Gendron, G.~Rousset, {\em et~al.}, ``A modal approach to optical
  gain compensation for the pyramid wavefront sensor,'' in {\em Adaptive
  {Optics} {Systems} {VI}},  L.~M. Close, L.~Schreiber, and D.~Schmidt, Eds.,
  {\bf 10703}, 653 -- 670, SPIE  (2018).
\newblock Backup Publisher: International Society for Optics and Photonics.

\bibitem{cavarroc_fundamental_2006}
C.~Cavarroc, A.~Boccaletti, P.~Baudoz, {\em et~al.}, ``Fundamental limitations
  on {Earth}-like planet detection with extremely large telescopes,'' {\em
  {A\&A}} {\bf 447}, 397--403  (2006).

\bibitem{cantalloube_f_wind-driven_2020}
{Cantalloube, F.}, {Farley, O. J. D.}, {Milli, J.}, {\em et~al.}, ``Wind-driven
  halo in high-contrast images - {I}. {Analysis} of the focal-plane images of
  {SPHERE},'' {\em A\&A} {\bf 638}, A98  (2020).

\bibitem{lovis_atmospheric_2017}
C.~Lovis, I.~Snellen, D.~Mouillet, {\em et~al.}, ``Atmospheric characterization
  of {Proxima} b by coupling the {SPHERE} high-contrast imager to the
  {ESPRESSO} spectrograph,'' {\em {A\&A}} {\bf 599}, A16  (2017).

\bibitem{verinaud05}
C.~{V{\'e}rinaud}, M.~{Le Louarn}, V.~{Korkiakoski}, {\em et~al.}, ``{Adaptive
  optics for high-contrast imaging: pyramid sensor versus spatially filtered
  Shack-Hartmann sensor},'' {\em {MNRAS}} {\bf 357}, L26--L30  (2005).

\bibitem{rigaut98}
F.~J. {Rigaut}, J.-P. {Veran}, and O.~{Lai}, ``{Analytical model for
  Shack-Hartmann-based adaptive optics systems},'' in {\em Adaptive Optical
  System Technologies},  D.~{Bonaccini} and R.~K. {Tyson}, Eds., {\em Society
  of Photo-Optical Instrumentation Engineers (SPIE) Conference Series} {\bf
  3353}, 1038--1048  (1998).

\bibitem{fried90}
D.~L. Fried, ``Time-delay-induced mean-square error in adaptive optics,'' {\em
  J. Opt. Soc. Am. A} {\bf 7}, 1224--1225  (1990).

\bibitem{hardy98}
J.~Hardy, {\em Adaptive Optics for Astronomical Telescopes}, Oxford series in
  optical and imaging sciences, Oxford University Press  (1998).

\bibitem{marois_effects_2003}
C.~Marois, R.~Doyon, D.~Nadeau, {\em et~al.}, ``Effects of {Quasi}-{Static}
  {Aberrations} in {Faint} {Companion} {Searches},'' {\em EAS Publications
  Series} {\bf 8}, 233--243  (2003).

\bibitem{milli_adaptive_2016}
J.~Milli, D.~Mawet, D.~Mouillet, {\em et~al.}, ``Adaptive optics in
  high-contrast imaging,'' {\em arXiv:1701.00836 [astro-ph]} {\bf 439}, 17--41
  (2016).
\newblock arXiv: 1701.00836.

\bibitem{goebel_measurements_2018}
S.~B. Goebel, O.~Guyon, D.~N.~B. Hall, {\em et~al.}, ``Measurements of
  {Speckle} {Lifetimes} in {Near}-infrared {Extreme} {Adaptive} {Optics}
  {Images} for {Optimizing} {Focal} {Plane} {Wavefront} {Control},'' {\em
  {PASP}} {\bf 130}, 104502  (2018).

\bibitem{macintosh_speckle_2005}
B.~Macintosh, L.~Poyneer, A.~Sivaramakrishnan, {\em et~al.}, ``{Speckle
  lifetimes in high-contrast adaptive optics},'' in {\em Astronomical Adaptive
  Optics Systems and Applications II},  R.~K. Tyson and M.~Lloyd-Hart, Eds.,
  {\bf 5903}, 170 -- 177, International Society for Optics and Photonics, SPIE
  (2005).

\bibitem{milli_speckle_2016}
J.~Milli, T.~Banas, D.~Mouillet, {\em et~al.}, ``{Speckle lifetime in XAO
  coronagraphic images: temporal evolution of SPHERE coronagraphic images},''
  in {\em Adaptive Optics Systems V},  E.~Marchetti, L.~M. Close, and J.-P.
  Véran, Eds.,  {\bf 9909}, 1455 -- 1472, International Society for Optics and
  Photonics, SPIE  (2016).

\bibitem{mawet_keck_2018}
D.~Mawet, C.~Z. Bond, J.~R. Delorme, {\em et~al.}, ``Keck {Planet} {Imager} and
  {Characterizer}: status update,'' in {\em Adaptive {Optics} {Systems} {VI}},
  L.~M. Close, L.~Schreiber, and D.~Schmidt, Eds., {\em Society of
  {Photo}-{Optical} {Instrumentation} {Engineers} ({SPIE}) {Conference}
  {Series}} {\bf 10703}, 1070306  (2018).

\bibitem{Males20}
J.~R. {Males}, L.~M. {Close}, O.~{Guyon}, {\em et~al.}, ``{MagAO-X first
  light},'' in {\em Society of Photo-Optical Instrumentation Engineers (SPIE)
  Conference Series},  {\em Society of Photo-Optical Instrumentation Engineers
  (SPIE) Conference Series} {\bf 11448}, 114484L  (2020).

\end{thebibliography}
\bibliographystyle{spiejour}   % makes bibtex use spiejour.bst

\listoffigures
\listoftables

\end{spacing}
\end{document}